\documentclass[preprint]{aastex631} % referee's format

\usepackage[normalem]{ulem}
\usepackage{color}

\usepackage{pifont}
\definecolor{mygreen}{HTML}{009900}
\definecolor{amber}{rgb}{1.0, 0.75, 0.0}
\newcommand{\cmark}{\textcolor{mygreen}{\Large\text{\ding{51}}}}
\newcommand{\xmark}{\textcolor{red}{\Large\text{\ding{55}}}}
\newcommand{\maybemark}{\textcolor{blue}{\Large\text{\ding{45}}}}

\shorttitle{Can a dark matter cusp hold a stellar core?}
\graphicspath{{./}{figures/}}
\begin{document}

\title{Can cuspy dark matter dominated halos hold cored stellar mass distributions? 
%\modified{Cored stellar mass distributions are often inconsistent with cuspy cold dark matter dominated haloes}\\
%\modified{The baryon mass is expected to trace dark matter in the center of low mass galaxies}\\
%\modified{Stellar cores in isolated ultra-low mass galaxies trace dark matter halo cores}\\
%\modified{A gateway to investigate the dark matter distribution in ultra-low mass galaxies using starlight}\\
%\modified{Constraining the dark matter distribution in ultra-low mass galaxies from deep photometry}\\
%\modified{chat\_gpt: Using the Eddington Inversion Method to Constrain the Dark Matter Distribution in Ultra-Low Mass Galaxies}
}

\author[0000-0003-1123-6003]{Jorge S\'anchez Almeida} \affil{Instituto de Astrof\'\i sica de Canarias, La Laguna, Tenerife, E-38200, Spain} \affil{Departamento de Astrof\'\i sica, Universidad de La Laguna}

\author[0000-0001-5848-0770]{Angel R. Plastino} \affil{CeBio y Departamento de Ciencias B\'asicas, \\ Universidad Nacional del Noroeste de la Prov. de Buenos Aires, \\ UNNOBA, CONICET, Roque Saenz Pe\~na 456, Junin, Argentina}

\author[0000-0001-8647-2874]{Ignacio Trujillo} \affil{Instituto de Astrof\'\i sica de Canarias, La Laguna, Tenerife, E-38200, Spain} \affil{Departamento de Astrof\'\i sica, Universidad de La Laguna}



\begin{abstract}
According to the current  concordance cosmological model, the dark matter (DM) particles are collision-less and produce self-gravitating structures with a central cusp which, generally, is not observed. The  observed density tends to a central plateau or core, explained within the cosmological model through the gravitational feedback of baryons on DM.  This mechanism  becomes inefficient when decreasing the galaxy stellar mass so that in the low-mass regime ($M_\star \ll 10^6\,{\rm M_\odot}$) the energy provided by the baryons is insufficient to modify cusps into cores.  Thus, if cores exist in these galaxies they have to reflect departures from the collision-less nature of DM. Measuring the DM mass distribution in these faint galaxies is extremely challenging, however,  their stellar mass distribution can be characterized through deep photometry. Here we provide a way of using only the stellar mass distribution to constrain the underlying DM distribution. The  so-called Eddington inversion method allows us to discard pairs of stellar distributions and DM potentials requiring (unphysical) negative distribution functions in the phase space. In particular, cored stellar density profiles are incompatible with the Navarro, Frenk, and White (NFW)  potential expected from collision-less DM if the velocity distribution is isotropic and the system spherically symmetric. Through a case-by-case analysis, we are able to relax these assumptions to consider anisotropic velocity distributions and systems which do not have exact cores. In general, stellar distributions with radially biased orbits are difficult to reconcile with NFW-like potentials, and cores in the baryon distribution tend to require cores in the DM distribution. 
\end{abstract}

\keywords{Cold dark matter (265), Dwarf galaxies (416), Galaxy mass distribution (606), Navarro-Frenk-White profile (1091), Galaxy dark matter halos (1880), Low surface brightness galaxies (940), Theoretical techniques (2093)}


%
%%%%%%%
%
\section{Introduction}\label{sec:intro}

The current concordance cosmological model assumes the dark matter particles to be cold and collision-less \citep[e.g.,][]{1978MNRAS.183..341W, 1984Natur.311..517B, 1985ApJ...292..371D, 1992ApJ...396L...1S, 2021arXiv210602672P,2022arXiv220307354B}. Thus, the cold dark matter (CDM) particles interact with themselves and with the baryons through gravitational forces only. Given the initial conditions set by the cosmological model, the CDM particles evolve under their own gravity to collapse into halos with {\em cusps} \citep[e.g.,][]{2014ApJ...790L..24C,2020MNRAS.495.4994B}, i.e., where the density is represented by the iconic NFW profile (Navarro, Frenk, and White \citeyear{1997ApJ...490..493N}) that grows boundlessly  when approaching the center of the gravitational potential.
This prediction contrasts with the fact that the observed dark matter (DM) haloes often show {\em cores}, i.e., their density tend to be constant as one approaches the center \citep[e.g.,][]{2015PNAS..11212249W,2017Galax...5...17D,2017ARA&A..55..343B}. This apparent contradiction is solved within the current CDM paradigm because the baryon dynamics  modifies the global gravitational potential also affecting the DM distribution and transforming the cusps into cores \citep[][]{1992Natur.356..489D,2010Natur.463..203G,2014MNRAS.437..415D}. This mechanism of feedback of baryons onto DM becomes inefficient when decreasing the galaxy mass, because the halo-to-stellar mass ratio increases with decreasing mass \citep[e.g.,][]{2013ApJ...770...57B}, reaching a point where the energy provided by star formation is simply not enough to modify the cusp of the CDM haloes \citep[e.g.,][]{2012ApJ...759L..42P,2015MNRAS.454.2092O}. The larger stellar mass unable to modify the inner slope of the DM profile is somewhat model dependent \citep[e.g.,][]{2016MNRAS.459.2573R}, but it roughly corresponds to stellar masses $M_\star < 10^{6}\,{\rm M}_{\odot}$ or halo masses $M_h < 10^{10}\,{\rm M_\odot}$ \cite[e.g.,][]{2014MNRAS.437..415D,2015MNRAS.454.2981C,2020ApJ...904...45H,2021MNRAS.502.4262J,2023MNRAS.519.4384E}. Thus, if galaxies with $M_h \ll 10^{10}\,{\rm M_\odot}$ show DM cores,  they are not due to baryon feedback but have to reflect the nature of DM:  whether it is fuzzy, self-interacting, warm, or else \citep[e.g.,][]{1994PhRvL..72...17D,2000PhRvL..85.1158H,2000PhRvL..84.3760S, 2022arXiv220307354B}.   

At these low masses, discerning observationally whether the DM halos have cores is extremely challenging, if not impossible. DM measurements require high spectral resolution spectroscopy to infer dynamical masses (whether optical, infrared, or radio wavelengths are used). The light is spread into small wavelength bins and so getting high signal-to-noise ratios is expensive observationally. On the contrary, stellar mass determinations depend on broad-band photometry, which is orders of magnitude faster than spectroscopy. Thus, measuring the baryon mass distribution in these low-mass objects is doable \citep[e.g.,][]{2021A&A...654A..40T} and, interestingly, low-mass galaxies tend to show cores in their stellar mass distribution \citep[e.g.,][]{2020ApJ...892...27M,2021ApJ...922..267C}. 
Since low-mass galaxies are often extremely DM dominated systems, one could naively think that the cores observed in stars just reflect the underlying DM mass distribution. If this conjecture turned out to be correct, it would provide a unique channel to study DM in low-mass galaxies, in a regime particularly informative to reveal the nature of DM  \citep[e.g.,][]{2015PNAS..11212249W,2017ARA&A..55..343B,2017Galax...5...17D}. Thus, the question arises as to whether the cores in the stellar mass distribution of DM dominated systems trace or not cores in the DM distribution.

%Even though it seems reasonable that the baryons follow, and so trace, the DM in these systems, we are not aware of previous works directly addressing the issue.
The purpose of this work is bringing up the question in the title to try to  give an answer in fairly broad terms. Thus, we show to be unlikely (although not impossible) that DM dominated systems with a central cusp have a stellar profile with a central core.  Therefore, our work provides a gateway to investigate the inner shape of the DM distribution in ultra-low mass galaxies using only their starlight.

We address the question using the so-called Eddington inversion method \citep{1916MNRAS..76..572E, 2008gady.book.....B, 2018JCAP...09..040L}. Simply put, it provides the distribution function (DF) in the phase space $f$ corresponding to a stellar mass density distribution $\rho$ immersed in a gravitational potential $\Phi$.  Given two arbitrary $\rho$ and $\Phi$, there is no guarantee that $f > 0$ everywhere, which is the absolutely minimum requirement for $\rho$ and $\Phi$ to be physically consistent. In this paper, we study the $f$ resulting from different combinations of $\rho$ (tracing the stars)  and $\Phi$ (dictated only by the  DM in ultra-low mass galaxies). We will show that unless the potential $\Phi$ is created by a mass distribution with a core, cored $\rho$s often give nonphysical $f < 0$. The computations in the paper neglect the contribution of the baryons to the overall potential, which we regard as a reasonable working hypothesis for the galaxies of interest. Thus, the gas in the ultra-low mass galaxies is not treated explicitly in the paper, but should play only a minor role in the analysis since it interacts with the stars only through its contribution to the gravitational potential. Therefore, as soon as the gas mass is much smaller than the total mass of the system, its presence can be neglected.

The paper is organized as follows: Sect.~\ref{sec:main_eqs} puts forward the Eddington inversion method together with  the main equations used in our analysis. The more lengthly derivations are separated in  Appendixes~\ref{app:appa} to \ref{app:alpha0}. Unphysical pairs $\rho$ -- $\Phi$ yielding  $f < 0$ somewhere are analyzed in Sect.~\ref{sec:positivity}. Among which one finds the prototypical cored $\rho$ immersed a NFW potential with isotropic velocities.
Examples and particular cases are worked out in Sect.~\ref{sec:results} to conclude that most often the cores in baryons trace DM cores in DM dominated self-gravitating systems. These results and their practical application to real galaxies are analyzed in Sect.~\ref{sec:conclusions}, including the effect of relaxing assumptions like spherical symmetry. 
Table~\ref{tab:summary} lists consistent and inconsistent combinations of $\rho$ and $\Phi$ resulting from our analysis.
In what follows, we use the terms baryons, stars, or particles indistinctly to refer to the component of the gravitationally bound system  that provides the density $\rho$.
  Moreover, in the context of this paper, the term low-mass galaxy is used to describe galaxies where the potential is approximately set by the DM because  the gravity produced  by the baryons can be neglected.

%
%%%%%%%%%%%%%%
%

%\section{Eddington, Cusps, and polytropes} %\bf EDDINGTON, CUSPS, AND POLYTROPES}

\section{The Eddington inversion method in our context}\label{sec:main_eqs}
This section provides a summary of the Eddington inversion method, and so, of the expressions used in Sects.~\ref{sec:positivity} and \ref{sec:results} to study whether cored baryon density distributions happen to be inconsistent with the gravitational potential created by CDM alone. We closely follow the approach and  terminology by \citet[][Sect.~4.3]{2008gady.book.....B}, but there are several alternative references on the subject \citep[e.g.,][]{1992MNRAS.255..561C,1996ApJ...471...68C,2018JCAP...09..040L}.
The main assumptions made when using the Eddington inversion method are \citep[][Sect.~4]{2008gady.book.....B}:  (1) the gravitational potential is smooth, (2) the trace particles (e.g., stars) have lifetimes larger than the crossing time, (3) the trace particles are collision-less, (4) the system is spherically symmetric, and (5) the system is described by a steady-state DF in the phase space. We take these assumptions as working hypotheses, which may not be fulfilled by particular objects but which may be good enough to describe large populations. For example, after a major merger the steady-state may require a few Gyr to be recovered \citep[e.g.,][]{2008MNRAS.391.1137L}, however, most galaxies only have a few such events during their lifetimes concentrated early on, therefore,  many galaxies should be in a quasi-steady state today. Based on these premises,  we first  consider particle systems with an isotropic velocity distribution.   Sect.~\ref{sec:eqs1} explains how to use the Eddington inversion method to recover the phase space DF from the three first spatial derivatives of the baryon density and of the gravitational potential.  The general expressions are particularized to specific mass distributions and gravitational potentials in Appendixes~\ref{sec:poly_nfw} and \ref{sec:plummer}.  Section~\ref{sec:eqs2} relaxes the assumption on the velocity isotropy, working out the expression of the DF for the Osipkov-Merritt velocity anisotropy model. Other anisotropic velocity models are considered too.  Even if contrived from a physical stand point, any gravitational potential is consistent with any density if the particles are arranged in perfectly circular orbits. The mixing model in Sect.~\ref{sec:mixing} describes the linear superposition of such a DF with circular orbits plus another DF with an isotropic velocity distribution. Finally, Sect.~\ref{sec:beta=const} treats the case of constant velocity anisotropy.

These physical systems and the corresponding DFs were chosen for simplicity, because they provide clear-cut constraints on the potential with relatively simple arguments.  There are extensions of the Eddington inversion method for other more general DFs that in principle could be used for similar diagnostics  \citep[e.g.,][]{1962MNRAS.123..447L,1987MNRAS.224...13D,1991MNRAS.253..414C,2017ApJ...838..123S}, but their study remains to be carried out, a task that requires specific follow-up work (Sect.~\ref{sec:conclusions}).

%
%%%%%
%
\subsection{Systems with isotropic velocity distribution}\label{sec:eqs1}

For spherically symmetric systems of particles with isotropic velocity distribution, the phase-space DF $f(\epsilon)$ depends only on the particle energy $\epsilon$. Then, the space density $\rho(r)$ turns out to be \citep[][Sect.~4.3]{2008gady.book.....B},
\begin{equation}
  \rho(r) = 4 \pi \sqrt{2} \,\int_0^{\Psi(r)} \, f(\epsilon) \sqrt{\Psi(r) - \epsilon} \, d\epsilon.
  \label{eq:leading}
\end{equation}
Here $\epsilon = \Psi - \frac{1}{2} v^2$ is the relative energy (per unit mass) of a  particle, and
$\Psi(r) = \Phi_0 - \Phi(r)$ is the relative potential energy, where
$\Phi(r)$ is the gravitational potential energy  and $\Phi_0$ is the gravitational
potential energy evaluated at the edge of the system.
For realistic systems, the relative potential $\Psi$ is a monotonically decreasing function
of the distance from the center $r$. Consequently, $\rho$ can be regarded as a function of $\Psi$. Differentiating $\rho$
with respect to $\Psi$,
\begin{equation} \label{derodepsi}
\frac{d \rho}{d \Psi} \, = \, 2\pi \sqrt{2}  \, \int_0^{\Psi} \, \frac{f(\epsilon)}{\sqrt{\Psi - \epsilon}}
\, d\epsilon.
\end{equation}
Inverting this Abel integral leads to Eddington's celebrated equation %\citep[e.g.,][Eq.~\[4.46\]]{2008gady.book.....B}
(e.g., \citealt{2008gady.book.....B}, Eq.~[4.46])
for the phase-space DF $f(\epsilon)$ in terms of the spatial density
$\rho(r)$,
\begin{equation}  \label{integralfe1}
f(\epsilon) = \frac{1}{2 \sqrt{2} \pi^2} \, \frac{d}{d \epsilon } \,
\int_0^{\epsilon} \, \frac{d\rho}{d\Psi} \, \frac{d\Psi}{\sqrt{\epsilon - \Psi}}.
\end{equation}
Integrating by parts twice,
\begin{equation} 
f(\epsilon) = \frac{1}{\sqrt{2} \pi^2} \,
\left[
\frac{1}{2 \sqrt{\epsilon}} \, \left(\frac{d\rho}{d\Psi} \right)_{\Psi=0} \,
+ \, \sqrt{\epsilon} \, \left(\frac{d^2\rho}{d\Psi^2} \right)_{\Psi=0} \, + \,
\int_0^{\epsilon} \, \frac{d^3\rho}{d\Psi^3} \, \sqrt{\epsilon - \Psi} \, d\Psi
\right].
\label{eq:boundary}
\end{equation}
The derivatives at the boundary, $\left(d\rho/d\Psi\right)_{\Psi=0}$ and $\left(d^2\rho/d\Psi^2\right)_{\Psi=0}$,  are in practice zero (see Appendix~\ref{app:boundary}), therefore,
\begin{equation} 
f(\epsilon) = \frac{1}{\sqrt{2} \pi^2} \,
\int_0^{\epsilon} \, \frac{d^3\rho}{d\Psi^3} \, \sqrt{\epsilon - \Psi} \, d\Psi.
\label{eq:ff1}
\end{equation}
To evaluate numerically the integral appearing in Eq.~(\ref{eq:ff1}), it is convenient to change the integration variable from $\Psi$ to $r$, because only $\rho(r)$ and $\Psi(r)$ are known explicitly. To use $r$ as integration variable, we need to express the derivatives of $\rho$ with respect to $\Psi$ in terms of the derivatives of $\rho$ and $\Psi$ with respect to $r$, i.e.,
\begin{equation} \label{drhodpsi1}
\frac{d\rho}{d\Psi} \, = \, \frac{d\rho/dr}{d\Psi/dr},
\end{equation}
\begin{equation} \label{drhodpsi2}
\frac{d^2\rho}{d\Psi^2} \, = \, \left( \frac{d\Psi}{dr} \right)^{-3} \,
\left[
\left( \frac{d^2\rho}{dr^2} \right) \left( \frac{d\Psi}{dr} \right)
-
 \left( \frac{d\rho}{dr} \right) \left( \frac{d^2\Psi}{dr^2} \right)
\right],
\end{equation}
\begin{equation} \label{drhodpsi3}
\frac{d^3\rho}{d\Psi^3} \, = \,
 \left( \frac{d^3\rho}{dr^3}  \right) \left( \frac{d\Psi}{dr} \right)^{-3} -
 3\left( \frac{d^2\rho}{dr^2}  \right)\left( \frac{d^2\Psi}{dr^2}  \right)
  \left( \frac{d\Psi}{dr} \right)^{-4}
  -
 \left( \frac{d\rho}{dr} \right) \left( \frac{d^3\Psi}{dr^3} \right)
  \left( \frac{d\Psi}{dr} \right)^{-4}
%  + 3 \left( \frac{d\rho}{dr} \right) \left( \frac{d^2\Psi}{dr^2}\right)^2 \left( \frac{d\Psi}{dr} \right)^{-5}
\end{equation}
\begin{displaymath}
  + 3 \left( \frac{d\rho}{dr} \right) \left( \frac{d^2\Psi}{dr^2}\right)^2 \left( \frac{d\Psi}{dr} \right)^{-5}.
\end{displaymath}
We now change the integration variable in the integral appearing in
Eq.~(\ref{eq:ff1}), %integralfe2}),}
\begin{equation} \label{integpsir}
\int_0^{\epsilon} \, \frac{d^3\rho}{d\Psi^3} \, \sqrt{\epsilon - \Psi} \, d\Psi
\, = \,
\int_{r_m}^{R} \, \frac{d\Psi}{dr} \frac{d^3\rho}{d\Psi^3} \, \sqrt{\epsilon - \Psi} \, dr
\, = \, - \,
\int_{R}^{r_m} \, \frac{d\Psi}{dr} \frac{d^3\rho}{d\Psi^3} \, \sqrt{\epsilon - \Psi} \, dr,
\end{equation}
where $R$ is the value of $r$ such that $\Psi(R) = \epsilon$, and $r_m$ is the maximum value of $r$, corresponding to the outer edge of the system. When the system has infinite spatial extent $r_m\to  \infty$.
Replacing  the expression (\ref{drhodpsi3}) for $d^3 \rho/ d\Psi^3 $
into the integral (\ref{integpsir}),

\begin{eqnarray} \label{integpsirnew}
\int_0^{\epsilon} \, \frac{d^3\rho}{d\Psi^3} \, \sqrt{\epsilon - \Psi} \, d\Psi
 &=& \int_{R}^{r_m} \, \left[
 - \left( \frac{d^3\rho}{dr^3}  \right) \left( \frac{d\Psi}{dr} \right)^{-2}
 +
 3\left( \frac{d^2\rho}{dr^2}  \right)\left( \frac{d^2\Psi}{dr^2}  \right)
  \left( \frac{d\Psi}{dr} \right)^{-3} \right. \cr
  &+& \left.
 \left( \frac{d\rho}{dr} \right) \left( \frac{d^3\Psi}{dr^3} \right)
  \left( \frac{d\Psi}{dr} \right)^{-3}
  -
 3 \left( \frac{d\rho}{dr} \right) \left( \frac{d^2\Psi}{dr^2}\right)^2
  \left( \frac{d\Psi}{dr} \right)^{-4} \right]
  \, \sqrt{\epsilon - \Psi} \, dr.
  \label{eq:ff2}
\end{eqnarray}

In short, according to Eqs.~(\ref{eq:ff1}) and (\ref{eq:ff2}), the DF $f(\epsilon)$ corresponding to a density $\rho(r)$ in a potential $\Phi(r)$ can be deduced from the first three derivatives of $\rho(r)$ and $\Psi(r)$ ($=\Phi_0-\Phi$). Appendix~\ref{app:appa} works them out in various practical cases involving polytropic $\rho$ and NFW densities and potentials. Examples of $\rho$ and $f(\epsilon)$ will be shown in Sect.~\ref{sec:results}, Figs.~\ref{fig:stability0_pub} -- \ref{fig:stability5a_pub}, \ref{fig:stability5b_pub}, and \ref{fig:stability5d_pub}.

%
%%%%%
\subsection{Systems with anisotropic velocity distribution: the Osipkov-Merritt model}\label{sec:eqs2}

The systems described in Sect.~\ref{sec:eqs1} have DFs depending on the particle energy only, which holds when the dispersion of velocities in the three independent spatial directions is the same. In terms of the so-called anisotropy parameter, these systems have $\beta(r) =0$, with
\begin{equation}
\beta (r) \, = \, 1 \, - \frac{\sigma_{\theta}^2 + \sigma_{\phi}^2}{2\sigma_r^2},
\label{eq:ani-param}
\end{equation}
where $\sigma_r$ is the radial velocity dispersion, and $\sigma_{\theta}$ and $\sigma_{\phi}$  are the tangential velocity dispersions in spherical coordinates.  The velocity isotropy requirement can be relaxed assuming  $f$ to depend not only on $\epsilon$ but also on the modulus of the angular momentum $L$. This is done in the  Osipkov-Merritt model, which  assumes a radial dependence of the anisotropy given by,
 \begin{equation} \label{betar}
   \beta(r) = \frac{r^2}{r^2 + r_b^2},
   \label{eq:ombeta}
 \end{equation}
 where the anisotropy radious $r_b$ sets the spatial scale of the changes in anisotropy. For $r \ll r_b$ the velocity distribution is isotropic, while for $r \gg r_b$ it is fully anisotropic with $\beta\to 1$ and the orbits becoming  mostly radial ($\sigma_{\theta}^2 + \sigma_{\phi}^2 \ll 2\sigma_r^2$). The assumption on $\beta(r)$ in the  Osipkov-Merritt model (Eq.~[\ref{eq:ombeta}]) may look artificial driven by analytical simplicity, but it is not quite so. This type of radial dependence of the anisotropy parameter is obtained in cosmological numerical simulations of galaxy formation in the low mass end of the mass spectrum \citep[e.g.,][]{2017ApJ...835..193E,2023arXiv230212818O}. In these simulations, the stars tend to have isotropic orbits in the center of the potential that turn into radial orbits in the outskirts. In the same numerical simulations, the DM haloes are more isotropic all over (discussed further in Sect.~\ref{sec:conclusions}).

Following \citet{2008gady.book.....B}, the phase-space DF of the Osipkov-Merritt model depends on the particle position and velocity through the quantity,
\begin{equation}
Q \, =\, \epsilon -  \frac{L^2}{2 r_b^2}.
\end{equation}
\noindent
It is convenient to define
\begin{equation}
\rho_{\rm OM}(r) \, = \, \left(1 + \frac{r^2}{r_b^2} \right) \,\rho(r).
\end{equation}
The connection between the  mass density and the phase space density can be expressed, in terms of $\rho_{\rm OM}$,
 in a way similar to the one corresponding to isotropic systems. Indeed, one has,
\begin{equation}
\rho_{\rm OM}(r) = 4 \pi \sqrt{2} \,\int_0^{\Psi(r)} \, f_{\rm OM}(Q) \sqrt{\Psi(r) - Q} \, dQ,
\end{equation}
\begin{equation} \label{derodepsiom}
\frac{d \rho_{\rm OM}}{d \Psi} \, = \, 2 \sqrt{2} \pi \, \int_0^{\Psi} \, \frac{f_{\rm OM}(Q)}{\sqrt{\Psi - Q}}
\, dQ,
\label{eq:osikov}
\end{equation}
\noindent
and
\begin{equation}
f_{\rm OM}(Q) = \frac{1}{2 \sqrt{2} \pi^2} \, \frac{d}{d Q} \,
\int_0^{Q} \, \frac{d\rho_{\rm OM}}{d\Psi} \, \frac{d\Psi}{\sqrt{Q - \Psi}}.
\end{equation}
Therefore, expressions (\ref{eq:ff1}) -- (\ref{eq:ff2}) also hold in this case replacing $\rho$ with $\rho_{\rm OM}$ and $\epsilon$ with $Q$.

%
%%%%
%
\subsection{Systems with anisotropic velocity distribution: the mixing model}\label{sec:mixing}

  A particle system having only circular orbits has always radial velocity equals zero, and so, $\sigma_r=0$, which leads to $\beta = -\infty$ everywhere. A system with such an extreme velocity anisotropy can reproduce any pair potential -- density with a DF, denoted here as $f_c$, guaranteed  to be positive everywhere (see Appendix~\ref{app:circular}).  A fairly general system with anisotropic velocity distribution can be constructed as a linear superposition of a system with circular orbits $f_c$ and a system with isotropic velocity distribution $f_i$ \citep{2008gady.book.....B}, so that
  \begin{equation}
      f({\bf r}, {\bf v}) = \mu\,f_i\left[\Psi (r)-v^2/2\right]+(1-\mu)\,f_c\left[{\bf r}, v_r,v_\theta,v_\phi\right],
  \label{eq:mixing}
\end{equation}
with ${\bf r}$ and ${\bf v}$ the position and velocity in the 6D phase space and $\mu$ parameterizing  the mixing fraction ($0\leq \mu\leq 1$). The symbols $ v_r,v_\theta,v_\phi$ represent the three coordinates of the velocity vector in a reference system where $v_r$ is the component in the radial direction set by ${\bf r}$. Equation~(\ref{eq:mixing}) explicitly shows that $f_i$ depends on the velocity through $v^2=v_r^2+v_\theta^2+v_\phi^2$, a property used in Sect.~\ref{sec:positivity} to discuss the feasibility of DFs from the mixing model. In this case, the anisotropy parameter at a fixed radius is 
\begin{equation}
\beta=-\frac{1-\mu}{\mu}\,\frac{\sigma_\theta^2+\sigma_\phi^2|_c}{\sigma_\theta^2+\sigma_\phi^2|_i} \leq  0,
\label{eq:beta2}
\end{equation}
where $\sigma_\theta$ and $\sigma_\phi$ stand for the velocity dispersion in the two tangential coordinates and $|_i$ and $|_c$ point out the isotropic and the circular velocity DF, respectively. The resulting orbits are between circularly biased and isotropic, but never radially biased.

%%%%
\subsection{Systems with anisotropic velocity distribution: constant velocity anisotropy}\label{sec:beta=const}
An extension of the above Eddington formalism deals with anisotropic velocity distributions of constant $\beta$. One starts from the DF,
\begin{equation}
  f(\epsilon, L) = L^{-2\beta} f_\epsilon(\epsilon),
  \label{eq:def33}
\end{equation}
which represents a leading order approximation for a wide class of DFs having the anisotropy parameter $\beta$ constant \citep[][]{2006ApJ...642..752A,2008gady.book.....B}. In these systems the DF depends not only on the relative energy $\epsilon$ but also on the modulus of the angular momentum $L$. Under this assumption, the mass volume density can be written as (\citealt{2008gady.book.....B}, Eq.~[4.66]),
\begin{equation}
  r^{2\beta}\rho(r) = \kappa_\beta \,\int_0^\Psi\frac{f_\epsilon(\epsilon)}{(\Psi-\epsilon)^{\beta-1/2}}\,d\epsilon,
  \label{eq:th1}
\end{equation}
where $\kappa_\beta$ is a positive numerical value independent of the radius $r$.  Note that this equation is formally quite similar to Eq.~(\ref{eq:leading}) provided $\beta < 1/2$, and so will be used in Sect.~\ref{sec:positivity} to point out the  inconsistency  of a large number of densities and potentials in way that parallels the isotropic case. This DF is also closely connected the so-called  {\em cusp slope-central anisotropy theorem} by \citet{2006ApJ...642..752A}, which links the inner slope of a density profile with the velocity anisotropy. It is examined in our context in Appendix~\ref{app:theorem}.

%
%%%%%%%
\section{Positivity of the phase-space distribution function}\label{sec:positivity}
 Positivity is the basic requirement for any physically sensible phase-space distribution.
 Given a relative potential function $\Psi$ and a mass density profile $\rho$,
 it is not guaranteed that the phase-space distribution yielded by the Eddington inversion
 method is positive everywhere in the phase space. A negative distribution function implies that the assumptions made  when applying Eddington's method are physically inconsistent: there is no phase-space DF that can reproduce the mass density $\rho$ under the assumed potential $\Psi$. We use this idea here and in Sect.~\ref{sec:results} to analyze the consistency of several combinations of $\Psi$ and $\rho$ that may be of practical importance.

Requiring $f$ to be non-negative constrains the properties of the centers of low-mass galaxies in fairly general terms. Equation~(\ref{derodepsi}) leads to a sufficient condition for the physical
 incompatibility between $\rho(r)$ and $\Psi(r)$  \citep[e.g.,][]{2018JCAP...09..040L}. If, for a given
 $\Psi $ (and, consequently, a given $r$), $d \rho / d \Psi$ vanishes, then, it
 follows from Eq.~(\ref{derodepsi})  that the phase-space density $f(\epsilon)$
 yielded by the Eddington method must reach negative values. (For the integral~[\ref{derodepsi}] to be zero with $f(\epsilon)\not=0$, $f(\epsilon) < 0$ somewhere within the interval $0\le \epsilon\le \Psi$.) Thus, if $d \rho / d \Psi=0$ somewhere, then no isotropic distribution is compatible with the given $\rho(r)$ and $\Psi(r)$. 
 Taking into account the relation,
 \begin{equation}
   \frac{d\rho}{d\Psi} \, = \, \frac{d\rho/dr}{d\Psi/dr},
   \label{eq:masterme}
 \end{equation}
  it follows that a cored mass density, defined as having
 \begin{equation}
  \lim_{r\to 0} \frac{d\rho}{dr} = 0,
   \label{eq:strict}
 \end{equation}
 is inconsistent with  a NFW background potential, which has
\begin{equation}
  \lim_{r\to 0}\frac{d\Psi}{dr} =-\frac{V_c}{2\,r_s^2}\ne 0;
  \label{eq:non-zero}
\end{equation}
 see Eq.~(\ref{eq:deriv_nfw}), with the constants $V_c$ and $r_s$ defined in Appendix~\ref{sec:poly_nfw}.

 The condition for $f>0$ derived above has a twist when the requirement of having a core (Eq.~[\ref{eq:strict}]) is somewhat relaxed. Consider a power law baryon density profile, $\rho\propto r^{-\alpha}$, with $\alpha=0$ for a cored profile. Consider also a power law for the density profile generating the potential,  $\rho_p\propto r^{-\alpha_p}$, with $\alpha_p=1$ for a NFW profile.  The relative potential $\Psi$ follows from $\rho_p$ so that for $\alpha\not= 0$ one finds\footnote{The relation is given explicitly in Sect.~\ref{sec:numerical},  Eqs.~(\ref{eq:esta}) and (\ref{eq:esta2}).},
%Then using Eqs.~(\ref{eq:esta}) and (\ref{eq:esta2}) to derived $\Psi$ from $\rho_p$,
\begin{equation}
  \frac{d\rho/dr}{d\Psi/dr}\simeq A\,\alpha\,\,r^{-(2+\alpha-\alpha_p)},
  \label{eq:ratio0}
\end{equation}
with $A> 0$ provided $0< \alpha_p < 3$. (The case $\alpha=0$ is controlled by a second term to be added to the RHS - right hand side -  of Eq.~[\ref{eq:ratio0}], and is treated in Appendix~\ref{app:alpha0}.) Thus the inconsistency between baryons and potential when $r\rightarrow 0$ disappears when $\alpha > 0$ since  $(d\rho/dr)\big/(d\Psi/dr)\rightarrow\infty$ when $r\rightarrow 0$. Note, however, that $f$ may still be negative somewhere else with $r\not=0$ even when $\alpha >0$, as we will show to be often the case (Sect.~\ref{sec:numerical}).  Note also that all profiles with $\alpha < 0$ (i.e., density decreasing toward the center) are discarded for any potential since the derivative Eq.~(\ref{eq:ratio0}) is either zero or negative when $r\rightarrow 0$. 

The above results hold for systems where the velocity anisotropy is zero, however, they can be extended to others more general anisotropic  systems. The  Osipkov-Merritt model, which has an anisotropy parameter given by Eq.~(\ref{eq:ombeta}), follows a relation for the DF (Eq.~[\ref{eq:osikov}]) formally identical to Eq.~(\ref{derodepsi}). Provided the density profile has a core (i.e., provided it follows Eq.~[\ref{eq:strict}]),
\begin{equation}
    \lim_{r\to 0}  \frac{d\rho_{OM}}{dr}=\lim_{r\to 0}\,\left(2r\rho/r_b^2+\left[1+r^2/r_b^2\right]\frac{d\rho}{dr}\right)=0,
  \end{equation}
which implies  that cored density profiles are incompatible with a NFW potential even when the velocity is anisotropic following an Osipkov-Merritt model. As we stress in Sect.~\ref{sec:eqs2}, this model for the radial variation of the anisotropy parameter is not as contrived as one may think since it is roughly followed by the low-mass model galaxies resulting from cosmological numerical simulations of galaxy formation.   

The constraints posed above happen to be a consequence of a more-general {\em  cusp slope-central anisotropy theorem} by \citet{2006ApJ...642..752A}. These authors showed that for systems with constant velocity anisotropy $\beta$ (i.e., those described in Sect.~\ref{sec:beta=const}), the need for the DF to be positive provides a constraint on the inner slope of the density profile $\alpha$ (i.e., $\rho\propto r^{-\alpha} {\rm ~when~} r\rightarrow 0$),
\begin{equation}
\alpha\geq 2\beta.
\end{equation}
This holds independently of the gravitational potential $\Psi$. As we show in Appendix~\ref{app:theorem}, when this is combined with the constraint in Eq.~(\ref{eq:non-zero}) set by having a NFW background potential, it leads to
\begin{equation}
  \alpha > 2\beta.
  \label{eq:inequality}
\end{equation}
The inequality (\ref{eq:inequality}) has a number of implications: (1) cores ($\alpha=0$) are inconsistent with isotropic velocities ($\beta=0$), as we have shown already, (2) cores  are inconsistent with radially biased orbits (i.e., only $\beta< 0$ is allowed for $\alpha=0$), (3) radially biased  orbits ($\beta > 0$) require cuspy baryon density profiles ($\alpha > 0$), and (3) circular orbits do not
%pose any problem since $\beta=-\infty$ for them.
impose any  restriction on the inner slope $\alpha$ since their $\beta = -\infty$.
Actually, it is already known that strongly tangentially biased orbits can reconcile a cored stellar density profile with a cuspy CDM-like background potential \citep[][a Schuster-Plummer density profile in a Hernquist potential]{2013A&A...558L...3B}.   

The constraint on galaxies having radially biased orbits ($\beta >0$) is particularly important from a practical point of view since these orbits seem to be the natural outcome of the formation of dwarf galaxies in $\Lambda$CDM  cosmological numerical simulations: see, e.g., \citet[][Fig.~2]{2017ApJ...835..193E} and \citet[][Fig.~5]{2023arXiv230212818O}. Moreover, even if the uncertainties are large, values  of $\beta \gtrsim 0$ are also observed among the DM dominated satellites of the MW \citep[e.g.,][]{2009MNRAS.394L.102L,2018NatAs...2..156M,2020A&A...633A..36M,2019MNRAS.484.1401R,2021MNRAS.500..410L,2022A&A...659A.119K}

There is also a fairly general family of tangentially biased DFs that can be discarded right away. It is described by the mixing model (Sect.~\ref{sec:mixing}) and covers the whole range of tangentially biased anisotropies from $\beta=-\infty $ to 0. The mixing model in Sect.~\ref{sec:mixing} combines circular orbit DFs ($f_c$ with $\beta = -\infty$) and isotropic velocity DFs ($f_i$ with $\beta = 0$) to produce tangentially biased DFs with $\beta < 0$ (Eq.~[\ref{eq:beta2}]). One may naively think that the always positive $f_c$ may compensate $f_i<0$ to yield a positive physically sensible DF $f =\mu\,f_i+(1-\mu)\,f_c$ (Eq.~[\ref{eq:mixing}]). However, all linear combinations  can be discarded for any $\mu \not=0$ if $f_i < 0$ somewhere. The argument goes as follows: assume that $f_i<0$ at ${\bf r}={\bf r_1}$ and ${\bf v}={\bf v_1}=(v_{r1},v_{\theta 1},v_{\phi 1})$ (see the dependencies of the DF on position ${\bf r}$ and velocity ${\bf v}$ in Eq.~[\ref{eq:mixing}]). Then  $f_i$ is also $<0$ at ${\bf r_1}$ and ${\bf v_2}=(v_{r2},0,0)$ provided $v^2_{r2}=v_{r1}^2+v_{\theta 1}^2+v_{\phi 1}^2$, since $f_i$ depends on ${\bf v}$ only through its modulus  $v$. However, $f_c({\bf r_1},v_{r2},0,0)=0$ because, by definition, $f_c$ only represents circular orbits that must have $v_r=0$. Thus, Eq.~(\ref{eq:mixing}) shows that $f({\bf r_1}, {\bf v_2}) <0$ and thus unphysical, with the only possible workaround of $\mu=0$, and so, of all orbits being circular.

On the basis of the above arguments, ultra-low mass galaxies for which the stellar mass distribution is well fitted with cored density profiles are dynamically incompatible with a NFW profile for the dark-mass component, at least if one assumes  that the phase-space DF of the stellar component depends only on the stellar energy,  or that is described by a Osipkov-Merritt model, or is anisotropic with radially biased orbits, or anisotropic with tangentially biased orbits following the mixing model. These  arguments cannot rule out a NFW background potential if other types of anisotropic phase-space distribution for the stars are assumed. For instance, a cored stellar profile may be compatible with a star distribution having a constant tangentially biased anisotropy (that is, having a constant and negative $\beta$).
To make the range of compatibilities more clear, Table~\ref{tab:summary} lists pairs of densities and potentials together with whether they  are consistent or inconsistent.

We note an important property of the consistency of a pair $\rho$~--~$\Psi$  based on whether $f(\epsilon) \ge 0$ $\forall \epsilon$.  If a particular pair is consistent or inconsistent, then any global factor affecting the density profile will not modify this character since $f(\epsilon)$ scales linearly with a multiplicative factor in $\rho$ (see  Eqs.~[\ref{eq:ff1}] and [\ref{eq:ff2}]). Thus, any of the inconsistencies brought out  here hold true independently of the (typically unknown) mass ratio between the stars and the DM halo creating the potential.

%%%%%%%
%
\section{Numerical results}\label{sec:results}
This section illustrates with specific examples the general results put forward in Sect.~\ref{sec:positivity}, analyzes the behavior outside the core of the system, and deals with profiles with shapes more complex than the ones considered in Sect.~\ref{sec:positivity}. We check  whether the DF resulting from particular pairs becomes negative at some point, which would discard the combination.  $\beta = 0$ is assumed, so the DF $f(\epsilon)$ follows from Eqs.~(\ref{eq:ff1}) and~(\ref{eq:ff2}). Equation~(\ref{eq:ff1}) is integrated numerically for every $\epsilon$ applying a Simpson's rule. The radial derivatives of $\rho$ and $\Psi$ in Eq.~(\ref{eq:ff2}) are computed analytically whenever possible using the equations in Appendix~\ref{app:appa}  (Sect.~\ref{sec:analytical}). Otherwise we compute them numerically (Sect.~\ref{sec:numerical}).

\begin{figure}[ht!]
\centering
\includegraphics[width=0.8\linewidth]{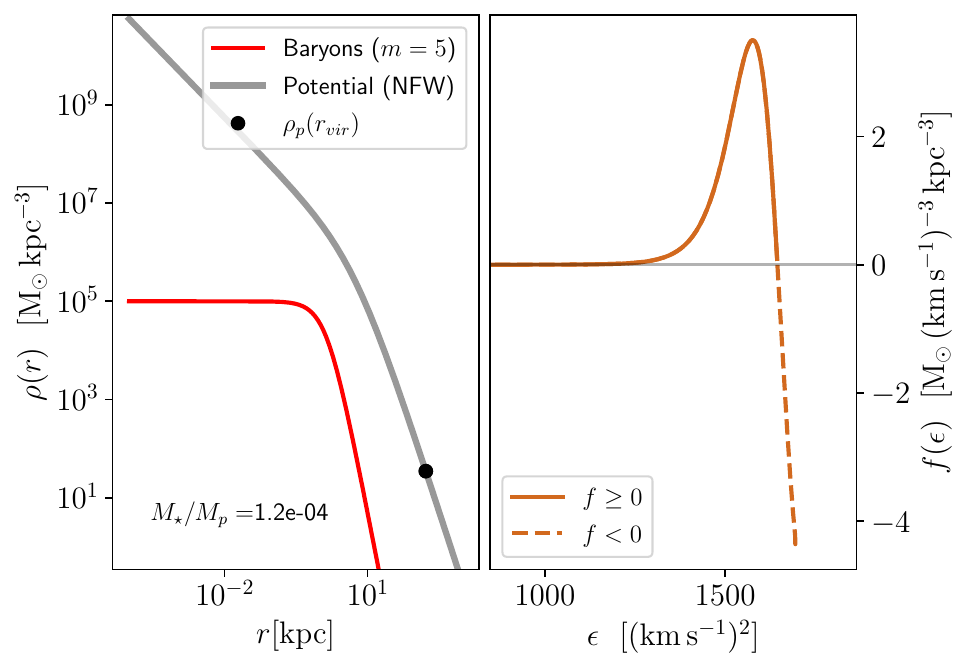}
\caption{
  Cored baryon distribution in a NFW gravitational potential. Left panel: a polytrope of order $m=5$  gives $\rho(r)$, with the central density $\rho(0)= 10^5\, {\rm M_\odot\, kpc^{-3}}$ and a core radius $r_0=1.4\,{\rm kpc}$, chosen so that $M_\star\sim 10^6\,{\rm M_\odot}$ (the red solid line). The NFW density profile that defines the overall gravitational potential through the Poisson equation (the gray line) has $\rho_s/\rho(0)=10$ and $r_s/r_0=4$, which provides $M_\star/M_{\rm p}\simeq 1.4\times10^{-4}$. The bullet symbol  points out the total density at the virial radius, assuming a concentration of 30 ($r_{vir}=30\,r_s$).  
Right panel: baryon DF $f$ needed to match baryon density and potential according to Eddington's inversion method which reaches negative values (the dashed line) implying that this particular combination  is unphysical.
}
\label{fig:stability0_pub}
\end{figure}
\subsection{Cored density in a NFW potential}\label{sec:analytical}
The computation of $f(\epsilon)$ is straightforward when  $\rho$ is a Schuster-Plummer profile,
\begin{equation}
  \rho(r)=\frac{\rho(0)}{\left[1+\left(r/r_0\right)^2\right]^{5/2}},
  \label{eq:sp_main}
\end{equation}
and $\Psi$ is described by a NFW potential (Appendix~\ref{app:appa}), a combination used here as reference of cored density profile immersed in a CDM-only potential.  The Schuster-Plummer profile is the polytrope of order $m=5$, and was chosen as reference because it provides a fair representation of the stellar mass distribution in real dwarf galaxies \citep[e.g.,][]{2021ApJ...921..125S}. As the rest of polytropes, this density profile has a core, therefore, it is not consistent with the potential derived from the cuspy NFW profile (Sect.~\ref{sec:positivity}).  An example is shown in Fig.~\ref{fig:stability0_pub}. The parameters that define this polytrope and the potential have been tuned to represent a realistic galaxy with stellar mass $M_\star\simeq 10^6\,{\rm M_\odot}$, core radius  $r_0=1.4~{\rm kpc}$, and total mass around $10^4$ times the stellar mass \citep[e.g.,][]{2013ApJ...770...57B,2016ApJ...817...84K}. The stars are immersed in the NFW potential generated by the matter distribution represented as the gray solid line in the left panel of Fig.~\ref{fig:stability0_pub}. (Note that this density fully defines $\Psi$ through Poisson's equation and independently of the velocity distribution of the particles creating the potential.) This component completely dominates the mass and the potential of the system: the mass within the gray solid line, $M_p$, has $M_p/M_\star\simeq 10^4$. As expected, $f< 0$ for some $\epsilon$ signaling that this combination of baryons and potential is unphysical.
%

%
%\begin{figure}[ht!] 
%\centering
%\includegraphics[width=0.8\linewidth]{stability2a_pub.pdf}
%\caption{
%  Cored density profile paired with the gravitational potential created by itself.  Left panel: the same density profile used in Fig.~\ref{fig:stability0_pub} (the red solid line), which is also the same mass distribution used to compute the potential (the gray solid line). Right panel: the corresponding DF which, as required for physical consistency, is positive for all $\epsilon$.  Integrated numerically to produce the figure, it  agrees with the analytic solution  in Eq.~(\ref{eq:theory}).
%}
%\label{fig:stability2a_pub}
%\end{figure}
%
For self gravitating systems, Poisson equation guarantees that  $f(\epsilon) \ge 0$ $\forall \epsilon$. For the stellar density profile shown in Fig.~\ref{fig:stability0_pub}, $f(\epsilon)$ is analytic (Eq.~[\ref{eq:theory}]). We use this fact to check the numerical integration scheme used to derive $f(\epsilon)$.
%An example is shown in Fig.~\ref{fig:stability2a_pub}, which corresponds to the same stellar density profile represented in Fig.~\ref{fig:stability0_pub} and its own gravity. In this case, $f(\epsilon)$ is analytic (Eq.~[\ref{eq:theory}]), and we checked that it coincides with the numerical integration used to produce the figure.  
%
%As we did in  Fig.~\ref{fig:stability0_pub}, rather than showing $\Psi$ in the example, we represent the density profile from which it stems, $\rho_p$, keeping in mind the one-to-one correspondence between a density profile and the potential it produces through Poisson equation.  

\subsection{Double power law density and potential}\label{sec:numerical}
\begin{figure}[ht!] 
\centering
\includegraphics[width=0.6\linewidth]{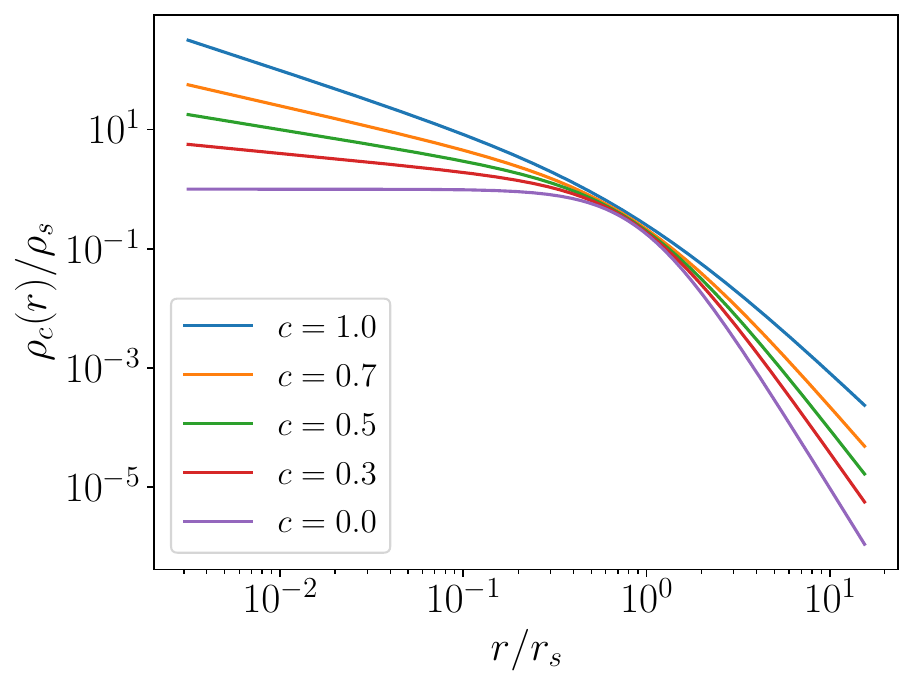}
\caption{
Doble power law density profile (Eq.~[\ref{eq:rhoc}]) that goes seamlessly from  a Schuster-Plummer profile to  a NFW profile  when the inner slope ($-c$) goes from 0 to -1. 
}
\label{fig:rhoc}
\end{figure}

A cored density in a NFW potential are inconsistent, as we have showed.
Here we expand the range of shapes to figure out how much the conditions for a core and a NFW potential can be relaxed and still getting inconsistent results. In our study, we use a family of density profiles commonly used in the literature \citep[e.g.,][]{1990ApJ...356..359H,2006AJ....132.2685M,2014MNRAS.441.2986D},
\begin{equation}
  \rho_{abc}(r) = \frac{\rho_s}{x^c(1+x^a)^{(b-c)/a}},
  \label{eq:numerical}
\end{equation}
with $x= r/r_s$, that encompasses both the NFW profile  ($a=1, b=3, {\rm ~and~} c=1$) and the  Schuster-Plummer profile ($a=2, b=5, {\rm ~and~} c=0$) shown in Fig.~\ref{fig:stability0_pub}.
%and \ref{fig:stability2a_pub}.
Actually, for $a=2$, $b=m$, and $c=0$,  $\rho_{abc}$ approximately accounts for the inner region of a polytrope of index $m$ \citep[e.g.,][]{2022Univ....8..214S} which is  important in this context since polytropes describe density profiles of self-gravitating N-body systems when reaching  thermodynamical equilibrium \citep[see,][]{1993PhLA..174..384P,2020A&A...642L..14S,2021MNRAS.504.2832S}.
The constants $r_s$ and $\rho_s$ in Eq.~(\ref{eq:numerical}) provide the global scaling for radius and density, respectively. The parameter $c$ gives  the inner logarithmic slope,
\begin{equation}
\lim_{r\to 0} \frac{d\log\rho_{abc}}{d\log r}=-c.
  \label{eq:central_slope}
\end{equation}
The three-parameter function $\rho_{abc}$ can be folded into a single parameter family  using $a=2-c$ and $b=5-2c$, 
\begin{equation}
  \rho_{c}(r) = \frac{\rho_s}{x^c(1+x^{2-c})^{(5-3c)/(2-c)}},
\label{eq:rhoc}
\end{equation}
which seamlessly scans from Schuster-Plummer to NFW when $c$ goes from 0 to 1 (see Fig.~\ref{fig:rhoc}).

In order to compute the DF of the baryons, one needs the derivatives of the density profile and the potential (Eqs.~[\ref{eq:ff1}] and [\ref{eq:ff2}]).  Using the Poisson equation for a spherically symmetric system \citep[e.g.,][]{2013MNRAS.428.2805A}, the potential and the required derivatives can be obtained in terms of the inner mass,
\begin{equation}
  M_p(<r) = 4\pi\,\int_0^r\,t^2\,\rho_p(t)\,dt,
  \label{eq:esta}
\end{equation}
so that 
\begin{equation}
  \Psi =\frac{G\,M_p(<r)}{r}+4\pi G\,\int_r^{\infty}\,t\,\rho_p(t)\,dt,
  \label{eq:pot_general}
\end{equation}
\begin{equation}
  \frac{d\Psi}{dr}= - \frac{G\,M_p(<r)}{r^2},
\label{eq:esta2}
\end{equation}
\begin{equation}
  \frac{d^2\Psi}{dr^2}=\frac{2GM_p(<r)}{r^3}-4\pi G\rho_p,
\end{equation}
and
\begin{equation}
  \frac{d^3\Psi}{dr^3}=-\frac{6GM_p(<r)}{r^4}+8\pi G \frac{\rho_p}{r}-4\pi G\frac{d\rho_p}{dr},
\label{eq:estaotra}
\end{equation}
which for  $\rho_p=\rho_{abc}$ can be computed analytically only for certain values of $a, b$ and $c$ \citep[][]{2013MNRAS.428.2805A}.

Employing, Eq.~(\ref{eq:numerical}) and Eqs.~(\ref{eq:esta}) -- (\ref{eq:estaotra}),  one can integrate numerically Eq.~(\ref{eq:ff2}) to obtain $f(\epsilon)$ via Eq.~(\ref{eq:ff1}). Using this approach, we  have scanned through a large number of pairs baryon densities and potentials both characterized by double exponential density profiles but with different parameters. Unless otherwise stated explicitly,  we employ the simplified version of the density given in  Eq.~(\ref{eq:rhoc}). The main results of our numerical exercise will be discussed next and are also summarized in Table~\ref{tab:summary}. To prevent confusion during the description, the parameters corresponding to the density profile that creates the potential are labeled with the subscript $p$ whereas those of the baryon density do not have any subscript. 
\begin{enumerate}
%
%%%%
  \item If the baryons have a core (i.e., if $c=0$) then the density generating the potential must also have a core to be consistent (i.e., $c_p=0$). This is shown by the numerical simulations (a counter-example with $c=0$, $c_p=0.05$ producing $f<0$ is shown in Fig.~\ref{fig:stability5a_pub}), but it also follows analytically from the study carried out in Appendix~\ref{app:alpha0} and discussed in the item \ref{item:this} below. The graphical summary in Fig.~\ref{fig:stability6a_plot} shows that when $c=0$,   $c_p$ must be zero for $f$ to be $>0$ everywhere. 
\begin{figure}[ht!] 
\centering
\includegraphics[width=0.6\linewidth]{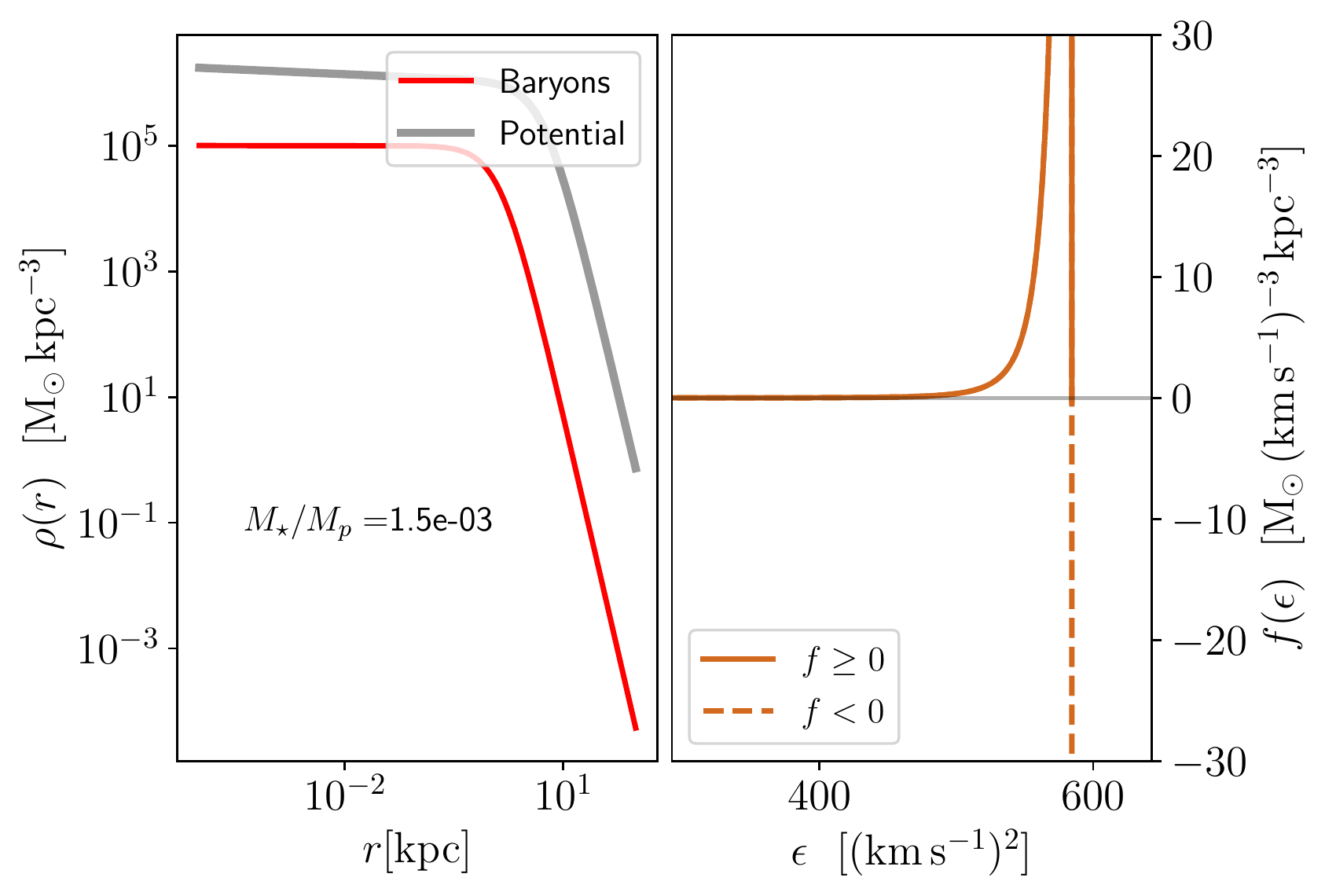}
\caption{Similar to Fig.~\ref{fig:stability0_pub} but this time only
with a hint of cusp in the density that generates the potential: $c=0$ and $c_p = 0.05$. Note that $f < 0$ at large $\epsilon$, implying  that density and potential are inconsistent with each other.  
}
\label{fig:stability5a_pub}
\end{figure}
\begin{figure}[ht!] 
\centering
\includegraphics[width=0.45\linewidth]{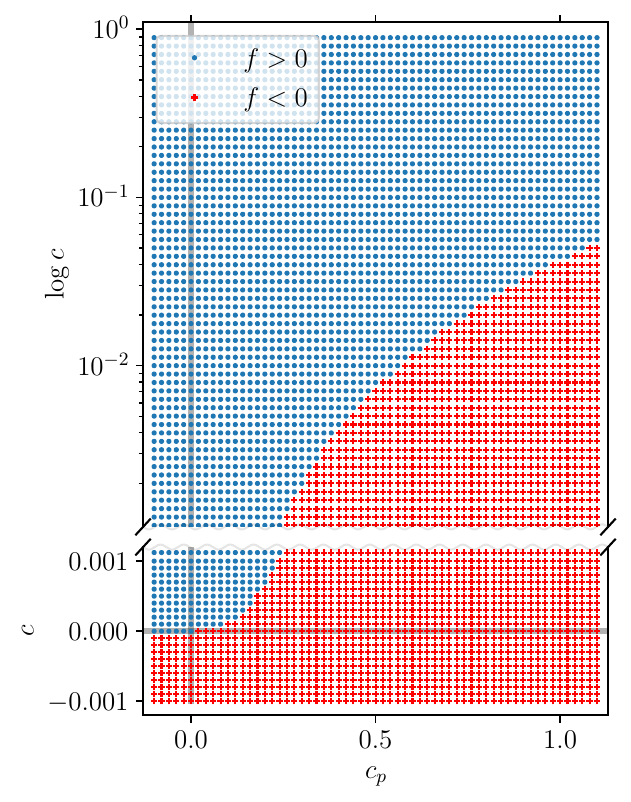}
\includegraphics[width=0.45\linewidth]{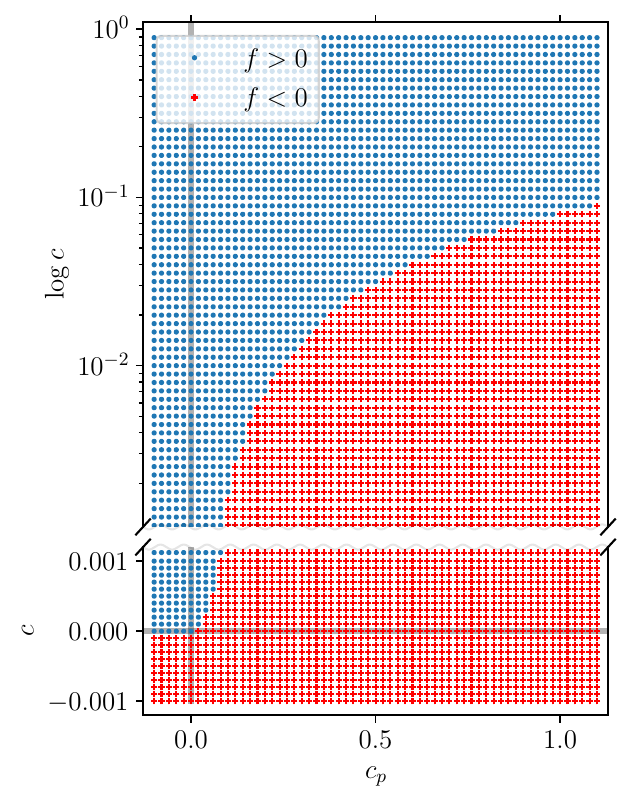}
\caption{
  Summary of the allowed (blue bullet symbols) and forbidden (red plus symbols) regions of the parameter space when both the baryon density profile and the density defining the underlying potential follow a profile described by Eq.~(\ref{eq:rhoc}) and  illustrated in Fig.~\ref{fig:rhoc}. The symbols $c$ and $c_p$ represent the inner slope of the density profile and the density generating the potential, respectively. The ordinate axis is split into two (logarithmic scale on top and linear scale at the bottom ) to show  the whole range of values of interest. The lines $x=0$ and $y=0$ are shown in light grey.  Left panel:  $r_s/r_{sp}=1/4$, with the baryons more centrally concentrated that the potential. Right panel:  $r_s/r_{sp}=2$, with the baryons more spread out than the potential. Baryons do not contribute to the overall potential and their velocity distribution is assumed to be isotropic ($\beta=0$) at all radii.
    In both cases, the sampling in $c_p$ is  $\Delta c_p=0.02$  whereas the sampling in $c$ is $\Delta c = 10^{-4}$ in the linear scale and $\Delta\log c=0.05$ in the logarithmic scale.
}
\label{fig:stability6a_plot}
\end{figure}
%
%%%
\item  If the baryon core is not perfect ($c \gtrsim 0$; denoted as soft-core in Table~\ref{tab:summary}), then a NFW profile may or may not be compatible with it.  Figure~\ref{fig:stability5b_pub} shows examples of incompatible (top panels) and compatible (bottom panels).  We have scanned a range of values for $c$ and $c_p$ ($-0.001 < c < 1$ and $-0.1 < c_p < 1.1$) with the compatibility summarized in Fig.~\ref{fig:stability6a_plot}. Roughly speaking, NFW potentials ($c_p=1$) are inconsistent with densities having $c\lesssim 0.1$.
\begin{figure}[ht!] 
\centering
\includegraphics[width=0.6\linewidth]{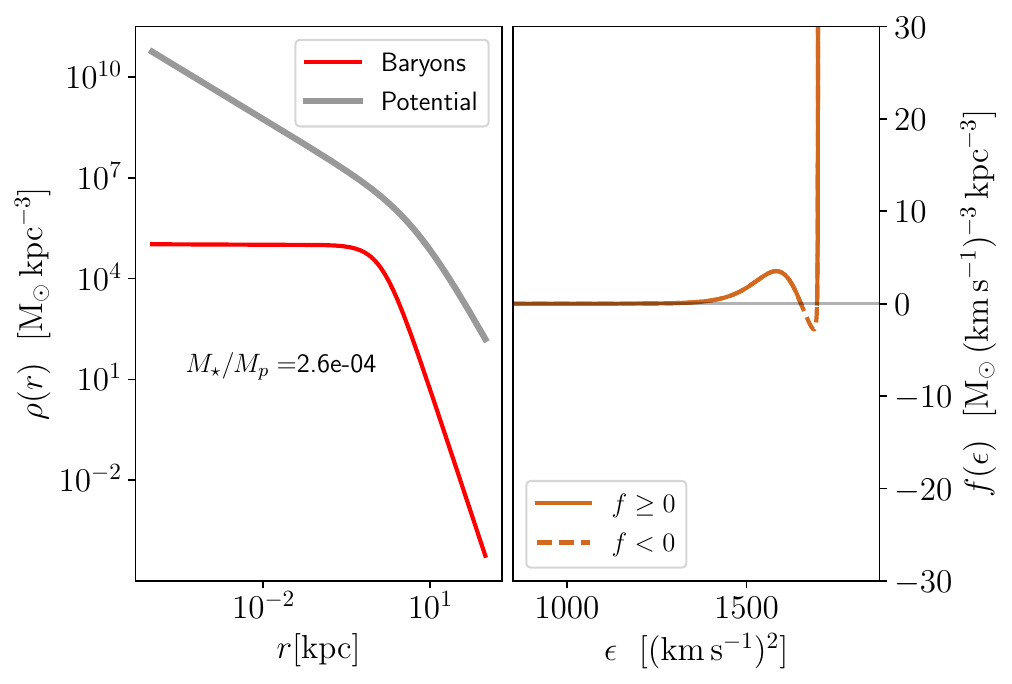}
\includegraphics[width=0.6\linewidth]{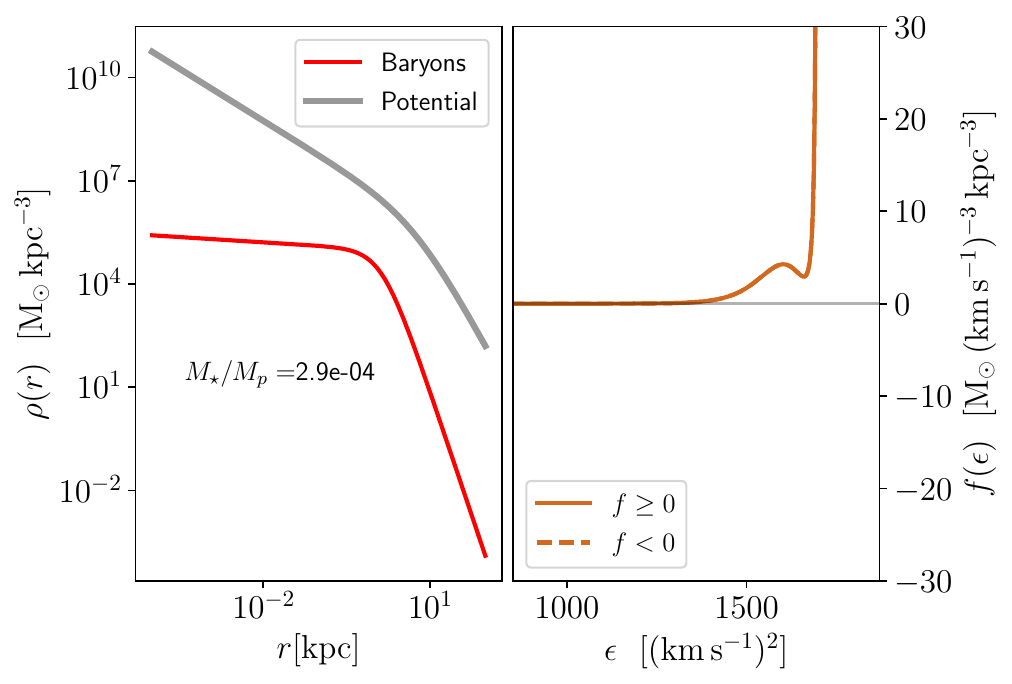}
\caption{
Top panels: similar to Fig.~\ref{fig:stability0_pub} but with a hint of cusp in the density: $c=0.005$ and $c_p = 1$. The DF $f < 0$ but not at the largest $\epsilon$. This combination of  baryon soft-core and NFW potential is still inconsistent. However, the inconsistency goes away as soon as the inner slope of the stellar density profile increases, as shown in the bottom panels, where   $c=0.1$ and $c_p = 1$. The global picture of compatibility -- incompatibility is summarized in Fig.~\ref{fig:stability6a_plot}. The shape of the profiles defining the baryon distribution and the potential is given by Eq.~(\ref{eq:rhoc}). 
}
\label{fig:stability5b_pub}
\end{figure}
\item Any density profile with $c=0$ and $a>2$ is physically irrealizable (see Fig.~\ref{fig:stability5d_pub}), independently of the potential.  The inconsistency remains even with the potential created by the self-gravity of the density, and means that no $\beta=0$ DF  is able to reproduce $a> 2$ profiles. The  behavior,  summarized in Fig.~\ref{fig:extrange}, is predicted analytically in Appendix~\ref{app:alpha0} and discussed further in item~\ref{item:this}.
\begin{figure}[ht!] 
\centering
\includegraphics[width=0.6\linewidth]{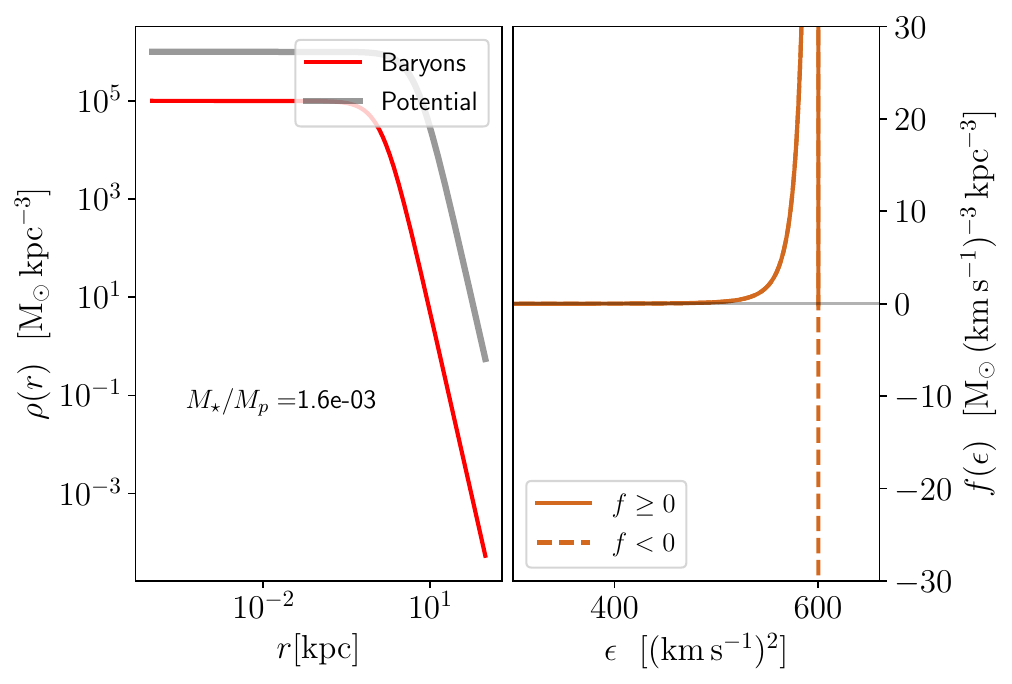}
\caption{
Similar to Fig.~\ref{fig:stability5b_pub} with  $c=c_p=0$ and $a=a_p=2.1$. Even if the two density profiles have the same shape,  $f < 0$ somewhere (the dashed line). This behavior for $a>2$ is predicted analytically in Appendix~\ref{app:alpha0} and discussed further in Sect~\ref{sec:numerical}, item~\ref{item:this}.
}
\label{fig:stability5d_pub}
\end{figure}
%%%
%
\begin{figure}[ht!] 
\centering
\includegraphics[width=0.49\linewidth]{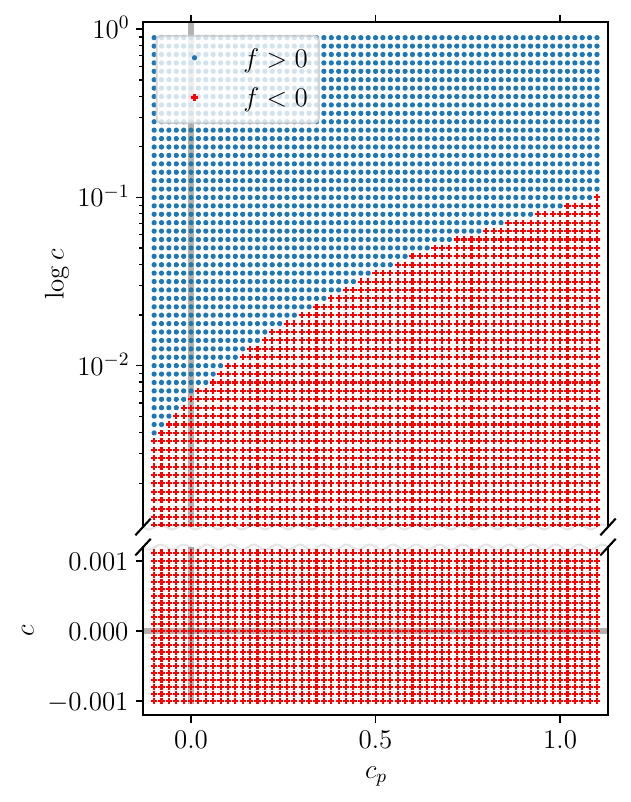}
\caption{
  Similar to Fig.~\ref{fig:stability6a_plot} but using $a=2.5-c$, so that Eq.~(\ref{eq:ratio2}) (or Eq.~[\ref{eq:ratio1}]) can be tested. Note that $c=c_p=0$ is unphysical even though $\rho$ and the density profile producing the potential are identical.  
}
\label{fig:extrange}
\end{figure}
\item A density profile significantly broader than the potential also yield inconsistent distribution functions. According to Fig.~\ref{fig:stability7b_plot},  $r_s/r_{sp} \lesssim 2$ for the DF to be non-negative, a constraint that may be used in real galaxies to set a lower limit to the size of the DM halo from the size of the observed starlight.  As we mention in Sect.~\ref{sec:positivity}, the density contrast between the density and the density producing the potential ($\rho_s/\rho_{sp}$) is irrelevant since it cannot change the sign of $f$.
\begin{figure}[ht!] 
\centering
\includegraphics[width=0.5\linewidth]{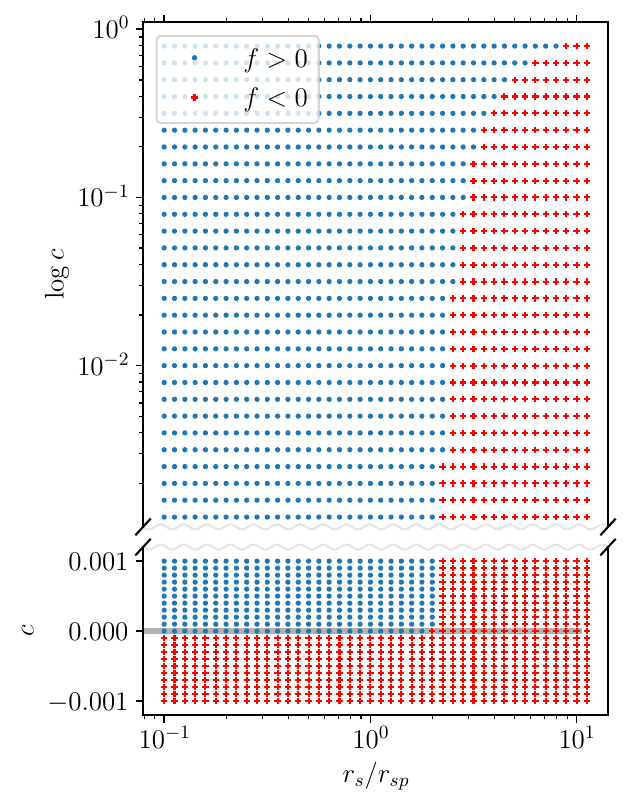}
\caption{
Diagnostic plot similar to Fig.~\ref{fig:stability6a_plot}, except that this time we represent simulations where both $c$ and $r_s/r_{sp}$ are varied. Note how $r_s$ cannot be larger than $\sim 2\,r_{sp}$ for $f$ to remain positive, a constraint that may be used  in real galaxies to set a lower limit to the size of the DM halo. 
In these simulations the $abc$ profile shapes of density and potential are identical, explicitly, $a=a_p$, $b=b_p$, and $c=c_p$, the three of them varying as for Eq.~(\ref{eq:rhoc}).
The sampling in $c$ is $\Delta c = 10^{-4}$ in the linear scale and $\Delta\log c=0.1$ in the logarithmic scale. The relative radii are shown in a logarithmic scale  with a sampling of $\Delta\log (r_s/r_{sp})=0.1$.
}
\label{fig:stability7b_plot}
\end{figure}
\item Note that the region where $c > c_p$, i.e., where the baryons are more cuspy than the halo,  presents no inconsistency in the summary plots of the  Figs.~\ref{fig:stability6a_plot}  and~\ref{fig:extrange}. We bring this fact up because some numerical simulations of ultra-low mass galaxies seem to show compact stellar concentrations having  $c > c_p\simeq 1$ \citep[e.g.,][]{2021MNRAS.504.3509O,2023arXiv230212818O}. These structures are physically feasible within the logical framework of our work which, among others, assumes negligible stellar mass ($M_\star\ll M_p$) and spherical symmetry.
  \item  \label{item:this} For the $abc$ densities that we are considering, the derivative used to diagnose the positivity of the DF in Sect.~\ref{sec:positivity} (Eq.~[\ref{eq:masterme}]) turns out to be (Eq.~[\ref{eq:ratio}], Appendix~\ref{app:alpha0}),
\begin{equation}
  \frac{d\rho/dr}{d\Psi/dr}\simeq \frac{D}{r^{2+c-c_p}}\,\Big[c+\frac{b-c}{r_s^a}\,r^a\Big],
 \label{eq:ratio2}
\end{equation}
with $D> 0$ for $c_p < 3$.  If $c\not=0$, the first term in the RHS of Eq.~(\ref{eq:ratio2}) dominates the behavior of the ratio when $r\rightarrow 0$. This term is identical to Eq.~(\ref{eq:ratio0}) with $\alpha=c$ and $\alpha_p=c_p$, and to allow for the derivative to differ from zero (and so for the DF to be positive), it only demands $c > c_p-2$. This is a very loose constraint and actually, most of the forbidden (red) region in Figs.~\ref{fig:stability6a_plot} and \ref{fig:extrange} actually meets this requirement. These two results, i.e., having $d\rho/d\Psi\not= 0$ and but $f<0$ somewhere, are consistent because the first one is more demanding than the second since it requires the (weighted) integral of $f$ to be positive (Eq.~[\ref{derodepsi}]), which can be met even when $f<0$ for some values of $\epsilon$ (see the example in the top  panel of Fig.~\ref{fig:stability5b_pub}).   
When $c=0$, the second term in the RHS of Eq.~(\ref{eq:ratio2}) rules, and then the potential and the density profiles would be inconsistent when $2-c_p-a > 0$ since the ratio of derivatives goes to zero. For $a=2$, as expected for polytropes, $c_p> 0$ is ruled out and the potential must have a core to be consistent with the core in the density profile. This condition for $c=0$ is truly restrictive, and is strictly followed by the simulations in Fig.~\ref{fig:stability6a_plot}. 
\end{enumerate}

%
%%%%%%%%
%
\section{Discussion  and Conclusions}\label{sec:conclusions}
%
%
%%%%%%%%% Summary Table
%
\begin{deluxetable*}{lccl}[h]
\tablecaption{Summary of the compatibility between baryon density profile ($\rho$) and  potential\label{tab:summary}}
\tablehead{
\colhead{Baryons \& Potential, Velocity} & 
\colhead{Consistency} &
\colhead{Comments}&
\colhead{Section}\\
\colhead{(1)}&
\colhead{(2)}&
\colhead{(3)}&
\colhead{(4)}
}
\startdata
Core\,$^\dagger$ \& NFW\,$^\ddag$, isotropic &\xmark& Eqs.~(\ref{eq:strict}) and (\ref{eq:non-zero}). $\beta=0\,^*$. Fig.~\ref{fig:stability0_pub}~~  & Sect.~\ref{sec:positivity}  \\
Power law\,$^\S$ \& Power law, isotropic&\maybemark& $\alpha > 0$\,$^\S$\,\cmark~$\alpha < 0$\,\xmark. Eq.~(\ref{eq:ratio0}). $\beta=0$  &Sects.~ \ref{sec:positivity}, \ref{sec:numerical}\\
Core \& Soft-core\,$^\#$, isotropic & \xmark&$\beta=0$. Fig.~\ref{fig:stability5a_pub}. Fig.~\ref{fig:stability6a_plot}&Sect.~\ref{sec:numerical}, App.~\ref{app:alpha0}\\
Core \& Core, isotropic & \maybemark&$\beta=0$. $a\leq 2$\,\cmark $a > 2$\,\xmark.
Fig.~\ref{fig:stability5d_pub} & Sects.~\ref{sec:analytical}, \ref{sec:numerical}, App.~\ref{app:alpha0}\\
Soft-core \& NFW, isotropic & \maybemark&$\beta=0$. Figs.~\ref{fig:stability6a_plot}, \ref{fig:stability5b_pub}. $c\gtrsim 0.1$\,\cmark $c \lesssim 0.1$\,\xmark. &Sects.~\ref{sec:positivity}, \ref{sec:numerical}\\
Soft-core \& Soft-core, isotropic & \maybemark&$\beta=0$.  Figs.~\ref{fig:stability6a_plot}, \ref{fig:stability5b_pub}&Sects.~\ref{sec:positivity}, \ref{sec:numerical}\\
&&$r_s \gtrsim 2\,r_{sp}$\,\xmark, $c> c_p$\,\cmark&Sect.~\ref{sec:numerical}\\
Core \& NFW, O-M model & \xmark& $\beta (\not= 0)$ in Eq.~(\ref{eq:ombeta})&Sect.~\ref{sec:positivity} \\
Core \& NFW, radially biased&\xmark&Constant $\beta$. $\beta > 0$&Sect.~\ref{sec:positivity}, App.~\ref{app:theorem}\\
Core \& Any, radially biased&\xmark&Constant $\beta$. $\beta > 0$&Sect.~\ref{sec:positivity}, App.~\ref{app:theorem}\\
Power-law \& Any, anisotropic & \maybemark &Constant $\beta$. $\alpha >  2\beta$&Sect.~\ref{sec:positivity},  App.~\ref{app:theorem}\\
Core \& NFW, circular&\cmark&$\beta= -\infty$&App.~\ref{app:circular}\\
Any \& Any, circular&\cmark&$\beta= -\infty$&App.~\ref{app:circular}\\
Any \& Any, tangentially biased &\maybemark&$\beta <0$. Eq.~(\ref{eq:mixing}). \xmark $f_i<0$ &Sects.~\ref{sec:mixing}, \ref{sec:positivity}\\
\enddata
\tablecomments{\\
  $^\dagger$ Core $\equiv\,d\log\rho/d\log r\to 0$ when $r\to 0$.\\
$^\ddag$ Navarro, Frenk, and  White potential (Eq.~[\ref{eq:nfwpotential}]) produced by a NFW profile (Eq.~[\ref{eq:nfw}]).\\
$^*$ Velocity anisotropy parameter $\beta$ defined in Eq.~(\ref{eq:ani-param}).\\
$^\S$ $\rho\propto r^{-\alpha}$.\\
  $^\#$ Soft-cores defined in Eqs.~(\ref{eq:numerical}) and (\ref{eq:rhoc}), and illustrated in Fig.~\ref{fig:rhoc}. Power laws $^\S$ are a particular type of those.\\
(1) Description of the baryon density, the gravitational potential, and the velocity distribution.\\
(2) The symbols \cmark , \xmark , and \maybemark\ stand for {\em compatible}, {\em incompatible}, and {\em may or may not}, respectively.\\ 
(3) Additional comments and keywords.\\
(4) Section of the text where the combination described in (1) is discussed.
}
\end{deluxetable*}
%
%%%%%%%%
% 
According to the current  concordance cosmological model, the DM particles are collision-less and, evolving under their own gravity, produce self-gravitating structures  that approximately  follow the iconic NFW profile with a cusp in its center (CDM haloes). These cusps ($\rho\propto r^{-1}$) are generally not observed in galaxies. The total density often presents a central plateau or core ($\rho\sim {\rm constant}$), which is believed to be produced by the coupling with baryons through gravity.  Star-formation driven outbursts modify the overall gravitational potential, affecting the CDM distribution too.  This mechanism of baryon feedback becomes inefficient when decreasing the galaxy stellar mass, reaching a point where the energy provided by baryons is simply not enough to modify the cusp of the CDM haloes  (see, Sect.~\ref{sec:intro} for references and details).  Despite all uncertainties and model dependencies, this threshold mass  roughly corresponds to isolated galaxies with stellar masses $< 10^{6}\,{\rm M}_{\odot}$ or halo masses $< 10^{10}\,{\rm M_\odot}$. Thus, if these ultra-low mass galaxies show cores,  they are not due to baryon feedback processes but have to reflect the nature of DM:  whether it is fuzzy, self-interacting, warm, or any of the other possibilities put forward in the literature.

Direct measurements of the DM mass distribution in these faint galaxies are difficult since they require high spectral resolution spectroscopy, which is observationally extremely challenging. However, there may be a  shortcut if the starlight somehow follows the DM since, even in low-mass low-luminosity galaxies,  deep photometry  is doable \citep[e.g.,][]{2021A&A...654A..40T}. One may naively think that stars must trace DM in these systems whose potential is fully dominated by DM. Nevertheless, stars are so weakly coupled with the DM that can potentially maintain a mass distribution differing from the DM distribution for longer than the age of the Universe \citep[e.g.,][]{2008gady.book.....B}. Thus, in order to use the {\em observable} stellar mass distribution as a proxy for the {\rm elusive} DM distribution, one has to show that somehow starlight traces DM in this DM dominated systems. More specifically, we know that low-mass galaxies often show cores in their stellar mass distribution (Sect.~\ref{sec:intro}). The question arises as whether this cored baryon distribution is or not consistent with the DM distribution expected from CDM particles (aka NFW profile). 
We address the question using the so-called Eddington inversion method. Under mildly restrictive assumptions (gravity from baryons negligible, stationary-state, smooth potential, and spherical symmetry; see, Sect.~\ref{sec:main_eqs}),  the method provides the DF in the phase space $f$ corresponding to a mass density distribution immersed in a gravitational potential.  Given two arbitrary density and potential, there is no guarantee that $f > 0$ everywhere, which is required for them to be physically consistent.

In this paper, we have studied different combinations of baryon density and gravitational potential that may help us to discern whether DM profiles in ultra-low mass galaxies have or not a core. We focus on the consistency of the various gravitational potentials with baryon density profiles showing a core (Eq.~[\ref{eq:strict}]) or soft-core (Eq.~[\ref{eq:central_slope}], with $c\gtrsim 0$).
The main conclusions of our analysis are summarized in Table~\ref{tab:summary} and can be expanded as follows:
\begin{itemize}
\item[-] Stellar cores in a NFW potential are incompatible  provided the velocity distribution is isotropic ($\beta=0$).
\item[-] Stellar cores and potentials stemming from a density with a quasi-core ($c_p >0$) are incompatible too. This result holds for isotropic velocities ($\beta =0$).
\item[-] As expected for physical consistency, stellar cores and potentials resulting from cored density profiles are consistent in isotropic ($\beta=0$) and radially biased systems ($\beta >0$).  
\item[-] Stellar cores and NFW potentials are also incompatible in systems with anisotropic velocities provided they follow the Osipkov-Merritt model. Even if artificial, it approximately describes the global trend expected in ultra-low mass galaxies, with $\beta\sim 0$ in the center and then increasing outwards ($\beta >0$).
\item[-]  Stellar cores and NFW potentials are incompatible in systems with radially biased orbits (${\rm constant~}\beta > 0$). Actually, a stellar core is incompatible with any potential without a core in systems with constant radially biased orbits ($\beta > 0$).
\item[-] Circular orbits ($\beta=-\infty$) can accommodate any combination of baryion density and potential, including a cored stellar density in a NFW potential. These configuration is very artificial, though. Unlikely to happen in real dwarf galaxies where orbits are expected to be radially biased (see the discussion below).
\item[-]
The linear superposition of two DFs is also a DF. Thus, one may think that the addition of a positive DF for circular orbits may compensate the negative DF for isotropic orbits to yield a positive physically sensible DF. However, this is not the case.  Independently of the relative weight, the mixing of an unphysical DF for isotropic velocities ($\beta=0$) with a physically realizable DF for circular orbits ($\beta =-\infty$) always yields unphysical DFs (Sect.~\ref{sec:mixing}).
\item[-] We denote as soft-cores those profiles where inner slope is not exactly zero but close to it ($c \gtrsim 0$). Soft-cores are inconsistent with NFW profiles when  $c \lesssim 0.1$ while they are consistent when  $c \gtrsim 0.1$. When the density profile that characterizes the potential also has a soft core (i.e., when $0\leq c_p\leq 1$), then the situation is more complicated as shown in, e.g., Fig.~\ref{fig:stability6a_plot}. This statements hold for isotropic velocity distributions.
\item[-] The inner slope of a soft stellar core and the radial anisotropy are related so that $c > 2\beta$.  In other words, large radially biased orbits are strongly inconsistent with soft stellar cores.  
\item[-] Positive inner slope in the stellar distribution, where the density grows outwards,  is discarded in every way.%\footnote{\modified{Curiosamente, this outward growing density is profile is predicted for baryon distribution in cored dwarfs within the MOND paradigm; see \citet{2022ApJ...940...46S}.}}. 
\item[-] For stellar densities and potentials with the same shape (whether cored or not), the stellar density distribution cannot be broader than twice the width of the density equivalent to the potential. This result refers to isotropic velocities and may be used in real galaxies to set a lower limit to the size of the DM halo from the size of the observed starlight.
\item[-] Pairs of density and potential where the inner slope of the density is larger than that of the potential ($c \geq c_p$) are not inconsistent.  This result refers to isotropic velocities too. 
\item[-] The above conclusions do not depend on a scaling factor on stellar density profile, therefore, they do not depend on the (unknown) ratio between the stellar mass and the total mass of the system.
\item[-] The functions used to represent the density and the potential are flexible enough to describe the central region in any polytrope of arbitrary index $m$ (Eq.~[\ref{eq:numerical}], with $a=2$, $b=m$, and $c=0$). Polytropes are important in the context of self-gravitating systems since they describe the density expected in N-body systems reaching thermodynamical equilibrium \citep[see,][]{1993PhLA..174..384P,2020A&A...642L..14S}. In other words, they portray the  DM density distribution  expected if the DM were not collision-less \citep[e.g.,][]{2021MNRAS.504.2832S}.  
\end{itemize}

How useful the above constraints are very much depends on the anisotropy  of the velocity field $\beta$ (Eq.~[\ref{eq:ani-param}]). In general, radially biased ($\beta >0$) and isotropic ($\beta=0$) orbits are more difficult to reconcile with a cuspy gravitational potential than tangentially biased orbits ($\beta < 0$).  The question arises as what is the anisotropy to be expected in real galaxies. This issue can be addressed from two complementary directions, namely, what is the anisotropy observed in the smallest galaxies, and what is the anisotropy recovered for the smallest galaxies formed in cosmological numerical simulations.
Even if the uncertainties are large because the estimates rely on measuring velocities of individual stars, the DM dominated satellites of the Milky Way (MW)  tend to have $\beta \gtrsim 0$ \citep[e.g.,][]{2009MNRAS.394L.102L,2018NatAs...2..156M,2020A&A...633A..36M,2019MNRAS.484.1401R,2021MNRAS.500..410L,2022A&A...659A.119K}. Note that these objects are not isolated galaxies and their internal baryon structure may be  strongly mediated by the presence of the MW and its circum-galactic medium through tidal forces, ram-pressure,  and starvation \citep[e.g.,][]{2004IAUS..217..440C,2010PhR...495...33B,2017ApJ...835..159S}. However, the observed trend is consistent with numerical simulations.  Radial anisotropies seems to be the natural outcome of the formation of dwarf galaxies in $\Lambda$CDM  cosmological numerical simulations: see, e.g., \citet[][Fig.~2]{2017ApJ...835..193E} and \citet[][Fig.~5]{2023arXiv230212818O}. Moreover, $\beta$ tends to zero when approaching the center of the gravitational potential, where the  stellar cores may be present and have to be observed. Thus,  $\beta \gtrsim 0$ at the centers seems to be a sensible conjecture when interpreting stellar mass distributions in real galaxies.

One of the seemingly more restrictive assumption leading to the constraints in Table~\ref{tab:summary} is the spherical symmetry of the density and potential. As it happens with the isotropy of the velocity field, the question of whether this is a good assumption for real ultra-low mass galaxies arises. Actually, the two issues are closely connected since, in real galaxies, both are set by the history of star-formation driven by cosmological gas accretion and mergers \citep[e.g.,][]{2009Natur.457..451D,2014A&ARv..22...71S}. In general, the smallest simulated galaxies tend to be rounded, although not perfectly spherical, with the DM component closer to  sphericity \citep[e.g.,][]{2002sgdh.conf..109B,2023arXiv230212818O}. On the other hand, the observed dwarf isolated galaxies are triaxial, but with three axes of similar lengths \citep[e.g,][]{2013MNRAS.436L.104R,2016ApJ...820...69S,2019ApJ...883...10P}. In addition to whether real ultra-low mass galaxies are or not well fitted by spherically symmetric models, independent  theoretical arguments point out that this assumption is not so critical since the incompatibilities may still hold when dropped. The extensions of the Eddington inversion method for axi-symmetric systems \citep[][]{1962MNRAS.123..447L,2008gady.book.....B} lead to expressions for the DF similar to Eqs.~(\ref{derodepsi}), (\ref{derodepsiom}), and (\ref{eq:th2}). They are expected to lead to restrictions similar to those worked out in this paper.   We are presently exploring them with promising results.
There are also extensions or variants of the Eddington inversion approach, suitable for
other more general spherically symmetric DFs, that in principle could be used for diagnostics and would be worth considering \citep[e.g.,][]{1987MNRAS.224...13D,1991MNRAS.253..414C,2017ApJ...838..123S}, but their  analysis remains to be carried out.

The constraints  in Table~\ref{tab:summary} result from treating particular cases, each one with its own peculiarities. The analysis of other cases (e.g., the study of axi-symmetric systems mentioned above) will enlarge a list which at present contains only a fraction of the constraints  yet to be discovered. In this sense, our work is only a pathfinder that shows how the traditional Eddington method can be used to study DM haloes in ultra-low mass galaxies.  Given the observed stellar distribution, the method seriously limits the properties of the DM halo where it resides.
Moreover, its interest probably exceeds the original scope that motivated the present study, and may be of application to other astrophysical systems where the stars represent only a minor fraction of the total mass, for example, the intra-cluster light as tracer of the DM galaxy cluster potential \citep[e.g.,][]{2019MNRAS.482.2838M,2022ApJ...940L..51M}.

In short, the question in the title of the paper, {\em Can CDM matter halos hold cored stellar mass distributions?}, has no simple {\em yes} or {\em no} answer. Instead, we find it to be unlikely, although not imposible,  than cored stellar mass distributions can be hosted in NFW DM haloes, provided the system is spherically symmetric. Thus, our work supports the interest of determining surface brightness profiles of ultra-low mass galaxies to constrain the nature of DM.  This work can be used as a guide to interpret observations so that the closer the observed galaxies to the hypotheses (spherically symmetry, stationarity, velocity isotropy, etc.) the more useful the constraints in Table~\ref{tab:summary}.

  Our ultimate goal is applying  the mathematical tools developed in this paper to observed dwarf galaxies with masses low enough  to constrain the nature of DM (Sect.~\ref{sec:intro}). This challenging task still requires several intermediate steps to be completed. In our roadmap,  we would like to test the machinery with the few local group galaxies for which independent information on the DM halo and on the stellar distribution is available \citep[e.g.,][]{2022NatAs...6..659B}, to see whether the constraints imposed by the Eddington inversion method and  by the kinematical measurements are consistent. We also need to know what is the signal-to-noise ratio and the number of targets required to make firm claims. Having a hundred targets with surface brightness profiles reaching down to 30\,mag\,arcsec$^{-2}$ seems to be doable \citep[e.g.,][]{2021ApJ...922..267C} but, does it suffice? Finally, we have to carefully select the actual data set of faint isolated dwarf galaxies. The two requirements are in tension since intrinsically faint galaxies are nearby and so tend to be satellites, but both are needed. One obvious possibility is waiting for better data \citep[e.g.,][]{2019ApJ...873..111I,2021ApJS..255...20A,2021A&A...654A..40T}. Alternatively, one can also think of studying the ultra-faint dwarfs of the local group \citep[e.g.,][]{2020ApJ...892...27M} cherry-picking those where the tidal forces and other environmental effects may be minimal (e.g., with large pericentic passage) and which truly proceed from low mass progenitors \citep[e.g.,][]{2003AJ....125.1926G}. Tidal forces change the internal structure of satellites and  reduce their stellar mass content, thus blurring any clear-cut interpretation of the observed DM distribution in terms of the DM nature, a caveat to keep in mind if this pathway is chosen.   All these works are currently ongoing or planed.

\begin{acknowledgments}
Thanks are due to 
Claudio Dalla-Vechhia for insightful discussions during the early stages of the work, and
to  Giuseppina Battaglia, Arianna Di cintio, and Ruben S\'anchez-Janssen for references.
Thanks are due to Matthew Orkney and Justin Read for discussions and clarifications on the velocity anisotropy and mass profile of the galaxies in their  simulations.     
JSA acknowledges financial support from the Spanish Ministry of Science and Innovation  (MICINN), project PID2019-107408GB-C43 (ESTALLIDOS). His visit to La Plata was partly covered by the MICINN through the Spanish State Research Agency, under Severo Ochoa Centers of Excellence Programme 2020-2023 (CEX2019-000920-S). JSA also wants to explicitly thank Angel Luis  Platino and the {\em Facultad de Ciencias Econ\'omicas de La Universidad Nacional de La Plata} for their hospitality during this visit. 
ARP acknowledges support to visit the IAC from the {\em Fundaci\'on Jes\'us Serra} and the IAC under their Visiting Researcher Programme 2020--2022. %
IT acknowledges support from the Project PCI2021-122072-2B, financed by MICIN/AEI/10.13039/501100011033, and the European Union NextGenerationEU/RTRP and the ACIISI, Consejer\'{i}a de Econom\'{i}a, Conocimiento y Empleo del Gobierno de Canarias and the European Regional Development Fund (ERDF) under grant with reference PROID2021010044 and from the State Research Agency (AEI-MCINN) of the Spanish Ministry of Science and Innovation under the grant PID2019-107427GB-C32 and IAC project P/302302, financed by the Ministry of Science and Innovation, through the State Budget and by the Canary Islands Department of Economy, Knowledge and Employment, through the Regional Budget of the Autonomous Community.
\end{acknowledgments}


\vspace{5mm}
%\facilities{This is a theoretical work}


\software{%astropy \citep{2013A&A...558A..33A,2018AJ....156..123A},
   numpy \citep{2020Natur.585..357H}, %
   scipy \citep{2020SciPy-NMeth}
   % Cloudy \citep{2013RMxAA..49..137F}, 
          %Source Extractor \citep{1996A&AS..117..393B}
          }



\appendix

%
%%%%%%
\section{Analytic derivatives of the density and the potential}\label{app:appa}

According to Eqs.~(\ref{eq:ff1}) and (\ref{eq:ff2}), the DF $f(\epsilon)$ corresponding to a density $\rho(r)$ in a potential $\Psi(r)$ can be deduced from the first three derivatives of $\rho(r)$ and $\Psi(r)$. This appendix works them out for various practical cases that involve polytropes and NFW potentials. They all are used in the main text.

\subsection{Distribution function for a Schuster-Plummer stellar mass density in a NFW potential}\label{sec:poly_nfw}

 The  Schuster-Plummer density (Eq.~[\ref{eq:sp_main}]) is defined as,
\begin{equation}
  D(r) = \rho(0) \left[1 + \frac{r^2}{r_0^2} \right]^{-\frac{5}{2}},
  \label{eq:sp_this}
\end{equation}
so that,
\begin{equation}
\frac{d D}{dr} = D_1(r)  =  -\frac{5\rho(0)}{r_0^2} \,
r \, \left[1 + \frac{r^2}{r_0^2} \right]^{-\frac{7}{2}},
\end{equation}
\begin{equation}
\frac{d^2 D}{dr^2} = D_2(r)  =  -\frac{5\rho(0)}{r_0^2}
 \left[1 + \frac{r^2}{r_0^2} \right]^{-\frac{7}{2}} +
 \frac{35 \rho(0)}{r_0^4} \, r^2
\, \left[1 + \frac{r^2}{r_0^2} \right]^{-\frac{9}{2}},
\end{equation}
and
\begin{equation}
\frac{d^3 D}{dr^3} = D_3(r)  =  \frac{35 \rho(0)}{r_0^4}\, r \,
 \left[1 + \frac{r^2}{r_0^2} \right]^{-\frac{9}{2}} +
 \frac{70 \rho(0)}{r_0^4}\, r \,
 \left[1 + \frac{r^2}{r_0^2} \right]^{-\frac{9}{2}} -
 \frac{315 \rho(0)}{r_0^6} \, r^3
\, \left[1 + \frac{r^2}{r_0^2} \right]^{-\frac{11}{2}}.
\end{equation}

 On the other hand, the NFW density profile is defined as,
  \begin{equation}
    \rho_{\rm NFW}(r)=\frac{\rho_s}{(r/r_s)(1+r/r_s)^2},
    \label{eq:nfw}
   \end{equation}
 with $r_s$ and $\rho_s$ two constants.  It creates a potential given by \citep[e.g.,][]{2008gady.book.....B}, 
\begin{equation}
  \Phi_{\rm NFW}(r) = - \frac{V_c}{r} \ln \left(1 + \frac{r}{r_s} \right), 
  \label{eq:nfwpotential}
\end{equation}
with $V_c=4\pi G\rho_s r_s^3$. Then the relative potential $\Psi(r)$, denoted for the NFW profile as $V(r)$, turns out to be,  
\begin{equation}
  V(r) = \Phi_{\rm NFW}(\infty)  - \Phi_{\rm NFW}(r)  =\frac{V_c}{r} \ln \left(1 + \frac{r}{r_s} \right),
  \label{eq:vvphi}
\end{equation}
with its derivatives given by,
\begin{equation}
\frac{dV}{dr} = V_1(r) =
\frac{V_c}{r^2} \left[ \frac{r}{r+r_s} - \ln \left(1 + \frac{r}{r_s} \right) \right],
\label{eq:deriv_nfw}
\end{equation}
\begin{equation}
\frac{d^2V}{dr^2} = V_2(r) =
- \frac{2 V_c}{r^3} \left[ \frac{r}{r+r_s} - \ln \left(1 + \frac{r}{r_s} \right) \right]
- \frac{V_c}{r (r + r_s )^2},
\end{equation}
and
\begin{equation}
\frac{d^3V}{dr^3} = V_3(r) =
\frac{6 V_c}{r^4} \left[ \frac{r}{r+r_s} - \ln \left(1 + \frac{r}{r_s} \right) \right]
+ \frac{2V_c}{r^2 (r + r_s )^2} + \frac{V_c(3r + r_s)}{r^2 (r + r_s )^3}.
\end{equation}
Using Eqs.~(\ref{eq:ff1}) and (\ref{eq:ff2}), the DF corresponding to a density given by Eq.~(\ref{eq:sp_this}) and a potential set by Eq.~(\ref{eq:vvphi}) turns out to be, 
\begin{equation}
f(\epsilon) =  \frac{1}{\pi^2 \sqrt{2}}
 \int_{R}^{\infty} \, dr \, \sqrt{\epsilon - V(r)} \,\,
\left[
\frac{- D_3}{V_1^2} + \frac{3 D_2 V_2}{V_1^3}
+ \frac{D_1 V_3}{V_1^3} - \frac{3 D_1 V_2^2}{V_1^4}
\right], 
\label{eq:stability0}
\end{equation}
with the limit $R$ implicitly defined as   $\epsilon = V(R)$.

%
%%%%%%%%%
\subsection{Distribution function for a Schuster-Plummer stellar mass density in a   Schuster-Plummer potential}\label{sec:plummer}

The gravitational potential corresponding to the mass density in Eq.~(\ref{eq:sp_this}) is \citep[e.g.,][]{2008gady.book.....B},

\begin{equation}
\Phi_{\rm SP}(r) = - W_c \left(1 + \frac{r^2}{r_0^2} \right)^{-1/2},
\end{equation}
with 
\begin{equation}
W_c = \left(\frac{4\pi}{3} \right) G r_0^2,
\end{equation}
so that the corresponding relative potential becomes,
\begin{equation}
  W(r) = \Phi_{\rm SP}(\infty)  - \Phi_{\rm SP}(r)  =  W_c \left(1 + \frac{r^2}{r_0^2} \right)^{-1/2},
  \label{eq:wwphi}
\end{equation}
with its derivatives given by,
\begin{equation}
\frac{dW}{dr} = W_1(r) =
- \frac{W_c}{r_0^2} r \left(1 + \frac{r^2}{r_0^2} \right)^{-3/2},
\end{equation}
\begin{equation}
\frac{d^2W}{dr^2} = W_2(r) =
- \frac{W_c}{r_0^2}  \left(1 + \frac{r^2}{r_0^2} \right)^{-3/2}
+ \frac{3W_c}{r_0^4} r^2  \left(1 + \frac{r^2}{r_0^2} \right)^{-5/2},
\end{equation}
and
\begin{equation}
\frac{d^3W}{dr^3} = W_3(r) =
\frac{9 W_c}{r_0^4} r \left(1 + \frac{r^2}{r_0^2} \right)^{-5/2}
- \frac{15W_c}{r_0^6} r^3  \left(1 + \frac{r^2}{r_0^2} \right)^{-7/2}.
\end{equation}

Using Eqs.~(\ref{eq:ff1}) and (\ref{eq:ff2}), the DF corresponding to a density given by Eq.~(\ref{eq:sp_this}) and a potential set by Eq.~(\ref{eq:wwphi}) turns out to be,
\begin{equation}
f(\epsilon) =  \frac{1}{\pi^2 \sqrt{2}}
 \int_{R}^{\infty} \, dr \, \sqrt{\epsilon - W(r)} \,\,
\left[
\frac{-D_3}{W_1^2} + \frac{3 D_2 W_2}{W_1^3}
+ \frac{D_1 W_3}{W_1^3} - \frac{3 D_1 W_2^2}{W_1^4}
\right],
\label{eq:stability2}
\end{equation}
with the radius $R$ implicitly defined as  $\epsilon = W(R)$. In the case of a self-gravitating system, so that the Schuster-Plummer potential is the one created by the Schuster-Plummer mass density, then Eq.~(\ref{eq:stability2}) can be integrated analytically to yield,
\begin{equation}
 f(\epsilon) = \frac{\rho(0)}{W_c^5}\frac{120}{(2\pi)^{3/2} \Gamma(9/2)} \,\epsilon^{7/2},
  \label{eq:theory}
\end{equation}
an expression used to check our numerical evaluations of $f(\epsilon)$.

\section{The terms at $\Psi=0$ in the Eddington inversion method}\label{app:boundary}
%\modified{Option A: The relative potential $\Psi$ generated  by a spherically symmetric system of finite total mass behaves as  $\Psi \propto r^{-1}$ for $r \to \infty$ (e.g., Eq.~[\ref{eq:pot_general}]). In order to have a finite total mass, the density of an object with infinite spatial extent has to go as  $\rho \propto r^{-\alpha}$, with $\alpha > 3$.  Combining the asymptotic behaviors of $\Psi$ and $\rho$, one finds that  $d\rho/d\Psi \propto r^{1-\alpha}$ and $d^2\rho/d\Psi^2 \propto r^{2-\alpha}$, with $\alpha > 3$, when $r \to \infty$. Consequently,   $\left( d\rho/d\Psi \right)_{\Psi = 0} = \left( d^2\rho/d\Psi^2 \right)_{\Psi = 0} = 0$. The above argument is not strictly valid  if $\Psi$ stands for the  NFW potential, because the potential does not correspond to a mass distribution with finite total mass, and it behaves as $\Psi \propto r^{-1} \, \ln r$ for $r\to \infty$. However, if $\rho$ describes an object of finite mass, one can show  that  $\left( d\rho/d\Psi \right)_{\Psi = 0} = \left( d^2\rho/d\Psi^2 \right)_{\Psi = 0} = 0$, in spite of the logarithmic factor appearing in the NFW potential.    }
The relative potential $\Psi$ generated  by a spherically symmetric system of finite total mass behaves as  $\Psi \propto r^{-1}$ for $r \to \infty$ (e.g., Eq.~[\ref{eq:pot_general}]). Consider objects where $\rho \propto r^{-b}$ for $r\to\infty$ (e.g., Eq.~[\ref{eq:numerical}]).   Combining the asymptotic behaviors of $\Psi$ and $\rho$, one finds that  $d\rho/d\Psi \propto r^{1-b}$ and $d^2\rho/d\Psi^2 \propto r^{2-b}$. Consequently,   $\left( d\rho/d\Psi \right)_{\Psi = 0} = \left( d^2\rho/d\Psi^2 \right)_{\Psi = 0} = 0$ provided $b > 2$. The above argument is not strictly valid  if $\Psi$ stands for the  NFW potential, because it does not correspond to a mass distribution with finite total mass, and it behaves as $\Psi \propto r^{-1} \, \ln r$ for $r\to \infty$. However, if $b > 2$, using Eq.~(\ref{drhodpsi2}) one can show  that  still $\left( d\rho/d\Psi \right)_{\Psi = 0} = \left( d^2\rho/d\Psi^2 \right)_{\Psi = 0} = 0$ when $r\to \infty$, in spite of the logarithmic factor appearing in the NFW potential.

%
%%%%%%%%%%%%%%%
\section{Distribution function of an spherically-symmetric system with circular orbits}\label{app:circular}
Any density $\rho(r)$ can be reproduced with a system of spherically-symmetric circular orbits \citep[][Sect. 4.3.2]{2008gady.book.....B}. By definition, their radial velocity is zero, $v_r=0$, and their tangential velocity is equal to the circular velocity, $v_t=v_c(r)$, with  
\begin{equation}
  v_c^2(r)=\frac{G\,M_p(<r)}{r}.
  \label{eq:co1}
\end{equation}
The symbol $M_p(<r)$ stands for the inner mass creating the potential $\Phi$. A general DF with the required properties is,
\begin{equation}
f_c(r,v_t,v_r) = F(r)\,\delta(v_t-v_c)\,\delta(v_r),
  \label{eq:co4}
\end{equation}
where $\delta$ represents a Dirac-delta function and $F$ is a function to be set by the density.  Since, $\rho$ is recovered from the integral of $f_c$ over all velocities,
\begin{equation}
  \rho(r) =2\pi\int\!\!\!\!\int f_c\, v_t\,dv_t\,dv_r ,
\end{equation}
then
\begin{equation}
  F(r)=\frac{\rho(r)}{2\pi\,v_c(r)},
  \label{eq:co5}
\end{equation}
which, together with Eq.~(\ref{eq:co1}), uniquely defines $F$ for any combination of $\rho$  and $\Phi$.
% Moreover, Eq.~(\ref{eq:co5}) guarantees $F>0$ which, together with Eq.~(\ref{eq:co4}), assures $f_c>0$ thus discarding obvious inconsistencies between density and gravitational potential.
Note that even if the DF in Eq.~(\ref{eq:co4})  is not explicitly written in terms of $\epsilon$ and $L$, it is straightforward to verify that it is a stationary DF.

\section{The theorem by An \& Evans in our context}\label{app:theorem}
Section~\ref{sec:beta=const} puts forward the DF,
\begin{equation}
  f(\epsilon, L) = L^{-2\beta} f_\epsilon(\epsilon),
  \label{eq:def33}
\end{equation}
which represents a leading order approximation for a wide class of DFs having the anisotropy parameter $\beta$ constant \citep[][]{2006ApJ...642..752A,2008gady.book.....B}. In this case the DF depends not only on the energy $\epsilon$ but also on the modulus of the angular momentum $L$.
Under this assumption, the mass volume density can be written as
(\citealt{2008gady.book.....B}, Eq.~[4.66]), 
\begin{equation}
  r^{2\beta}\rho(r) = \kappa_\beta \,\int_0^\Psi\frac{f_\epsilon(\epsilon)}{(\Psi-\epsilon)^{\beta-1/2}}\,d\epsilon,
  \label{eq:th1}
\end{equation}
where $\kappa_\beta$ is a positive numerical value independent of the radius $r$. 
As we argued in Sect.~\ref{sec:positivity}, for the integral in the RHS of Eq.~(\ref{eq:th1}) to be zero, $f_\epsilon < 0$ somewhere which through Eq.~(\ref{eq:def33}) makes $f$ unphysical. Assuming   $\rho(r)\propto r^{-\alpha}$ when $r\rightarrow 0$, then the left-hand-side of Eq.~(\ref{eq:th1}) differs from zero if $2\beta-\alpha \leq 0$, which is the theorem proved by   \citeauthor{2006ApJ...642..752A}. Here we go a step further and provided $\beta <1/2$ (Eq.~[\ref{eq:th1}] diverges when taking derivatives and $\beta > 1/2$), one obtains  (\citealt{2008gady.book.....B} Eq.~[4.67]), 
\begin{equation}
  %\frac{d}{d\Psi}  \big[r^{2\beta}\rho(r)\big] =
\frac{d\big[r^{2\beta}\rho(r)\big]\big/dr}{d\Psi/dr}   = \kappa_\beta \,\left(\frac{1}{2}-\beta\right)\int_0^\Psi\frac{f_\epsilon(\epsilon)}{(\Psi-\epsilon)^{\beta+1/2}}\,d\epsilon.
  \label{eq:th2}
\end{equation}
In the case of $\Psi$ given by a NFW potential, its radial derivative at $r=0$ differs from zero and is negative (Eq.~[\ref{eq:non-zero}]) therefore, to avoid the RHS of Eq.~(\ref{eq:th2}) to be less or equal to zero (and so to avoid an unphysical $f_\epsilon<0$), $2\beta-\alpha\not= 0$ and $2\beta-\alpha-1 \leq 0$. The 2nd condition  is automatically met because   $2\beta-\alpha$ is already $\leq 0$ according to \citeauthor{2006ApJ...642..752A}. Together with this inequality, the first condition implies that for $f>0$ then
\begin{equation}
  \alpha > 2\beta.
\label{eq:ineq}
\end{equation}
Our derivation assumes $\beta < 1/2$, however, it is not difficult to show that the inequality still holds in the limit case when $\beta=1/2$ and Eq.~(\ref{eq:th2}) is not valid. % see 4.68 of B&T 1st ed, and ask f1 to be non-negative. 

There are several obvious consequences of the inequality in Eq.~(\ref{eq:ineq}): (1) cores ($\alpha=0$) are inconsistent with isotropic velocities ($\beta=0$), (2) cores  are inconsistent with radially biased velocities (i.e., only $\beta< 0$ is allowed), (3) radially biased  orbits ($0< \beta < 1/2$) require cuspy baryon density profiles ($\alpha > 0$), and (3) circular orbits do not pose any problem since $\beta=-\infty$.

%
%%%%%%%%%%%%%%%%%%%%%
\section{Value of $(\lowercase{d\rho/dr)\big/(d\Psi/dr)}$ when $\lowercase{r\rightarrow 0}$ and $\lowercase{d\rho/dr \rightarrow 0}$}\label{app:alpha0}

Starting out from the definition of $\rho_{abc}$ in Eq.~(\ref{eq:numerical}), one finds for $r\rightarrow 0$,
\begin{equation}
  \frac{d\rho_{abc}}{dr}\simeq -\frac{\rho_s}{r_s}\,\frac{c+(b-c)\,x^a}{x^{1+c}},
\end{equation}
with $x=r/r_s$. Similarly, Eq.~(\ref{eq:esta2}) provides the 1st order approximation,
\begin{equation}
  \frac{d\Psi}{dr}\simeq -B\,r^{1-c_p},
\end{equation}
where $c_p$ is the value of $c$ of the density profile assumed to generate the potential $\Psi$ and $B$ is a positive constant. Putting together the two previous equations, one finds,
\begin{equation}
  \frac{d\rho/dr}{d\Psi/dr}\simeq \frac{D}{r^{2+c-c_p}}\,\Big[c+\frac{b-c}{r_s^a}\,r^a\Big],
 \label{eq:ratio}
\end{equation}
with $D> 0$ for $c_p < 3$.
In the range of interest for galaxies, the parameters $c$ and $c_p$ are from $\sim 0$ to $\sim 1$, with $c \leq c_p$, whereas $a$ is between $\sim 1$ and $\sim 2$. If $c\not=0$, the first term in the RHS of Eq.~(\ref{eq:ratio}) dominates the behavior of the ratio when $r\rightarrow 0$. This term is identical to Eq.~(\ref{eq:ratio0}) with $\alpha=c$ and $\alpha_p=c_p$, and its behavior is discussed in detail in Sect.~\ref{sec:positivity}.   The case when $c=0$ is particularly interesting since it represents a cored density profile. Only the 2nd term in the RHS of Eq.~(\ref{eq:ratio}) is not zero and it turns out to be,
\begin{equation}
  \frac{d\rho/dr}{d\Psi/dr}\simeq \frac{D\, b}{r_s^a}\frac{1}{r^{2-c_p-a}}.
 \label{eq:ratio1}
\end{equation}
Thus, the potential and the density profiles would be inconsistent when $2-c_p-a < 0$ since the ratio of derivatives goes to zero when $r\rightarrow 0$.  It implies that when  $a=2$ (e.g., Schuster-Plummer profile; Appendix~\ref{sec:plummer}), all $c_p > 0$ are physically irrealizable (see Fig.~\ref{fig:stability6a_plot} for $c=0$). It also implies that  when $c_p=0$, and so the potential has the same core as the density, the pair density and potential are unphysical for $a>2$. This somewhat surprising behavior has been checked numerically (Figs.~\ref{fig:stability5d_pub} and \ref{fig:extrange}).


\bibliography{biblio_paper142}{}
\bibliographystyle{aasjournal}



\end{document}